\documentclass[%
 reprint,
superscriptaddress,
 amsmath,amssymb,
 aps,
]{revtex4-1}

\usepackage{subcaption}
\usepackage{url}
\usepackage{color}
\usepackage{graphicx}% Include figure files
\usepackage{dcolumn}% Align table columns on decimal point
\usepackage{bm}% bold math
\begin{document}

\preprint{APS/123-QED}

\title{Measurement of the $\nu_{\mu}$ charged current quasi-elastic cross-section\\
on carbon with the T2K on-axis neutrino beam}% Force line breaks with \\

\newcommand{\INSTC}{\affiliation{University of Alberta, Centre for Particle Physics, Department of Physics, Edmonton, Alberta, Canada}}
\newcommand{\INSTEE}{\affiliation{University of Bern, Albert Einstein Center for Fundamental Physics, Laboratory for High Energy Physics (LHEP), Bern, Switzerland}}
\newcommand{\INSTFE}{\affiliation{Boston University, Department of Physics, Boston, Massachusetts, U.S.A.}}
\newcommand{\INSTD}{\affiliation{University of British Columbia, Department of Physics and Astronomy, Vancouver, British Columbia, Canada}}
\newcommand{\INSTGA}{\affiliation{University of California, Irvine, Department of Physics and Astronomy, Irvine, California, U.S.A.}}
\newcommand{\INSTI}{\affiliation{IRFU, CEA Saclay, Gif-sur-Yvette, France}}
\newcommand{\INSTGB}{\affiliation{University of Colorado at Boulder, Department of Physics, Boulder, Colorado, U.S.A.}}
\newcommand{\INSTFG}{\affiliation{Colorado State University, Department of Physics, Fort Collins, Colorado, U.S.A.}}
\newcommand{\INSTFH}{\affiliation{Duke University, Department of Physics, Durham, North Carolina, U.S.A.}}
\newcommand{\INSTBA}{\affiliation{Ecole Polytechnique, IN2P3-CNRS, Laboratoire Leprince-Ringuet, Palaiseau, France }}
\newcommand{\INSTEF}{\affiliation{ETH Zurich, Institute for Particle Physics, Zurich, Switzerland}}
\newcommand{\INSTEG}{\affiliation{University of Geneva, Section de Physique, DPNC, Geneva, Switzerland}}
\newcommand{\INSTDG}{\affiliation{H. Niewodniczanski Institute of Nuclear Physics PAN, Cracow, Poland}}
\newcommand{\INSTCB}{\affiliation{High Energy Accelerator Research Organization (KEK), Tsukuba, Ibaraki, Japan}}
\newcommand{\INSTED}{\affiliation{Institut de Fisica d'Altes Energies (IFAE), Bellaterra (Barcelona), Spain}}
\newcommand{\INSTEC}{\affiliation{IFIC (CSIC \& University of Valencia), Valencia, Spain}}
\newcommand{\INSTEI}{\affiliation{Imperial College London, Department of Physics, London, United Kingdom}}
\newcommand{\INSTGF}{\affiliation{INFN Sezione di Bari and Universit\`a e Politecnico di Bari, Dipartimento Interuniversitario di Fisica, Bari, Italy}}
\newcommand{\INSTBE}{\affiliation{INFN Sezione di Napoli and Universit\`a di Napoli, Dipartimento di Fisica, Napoli, Italy}}
\newcommand{\INSTBF}{\affiliation{INFN Sezione di Padova and Universit\`a di Padova, Dipartimento di Fisica, Padova, Italy}}
\newcommand{\INSTBD}{\affiliation{INFN Sezione di Roma and Universit\`a di Roma ``La Sapienza'', Roma, Italy}}
\newcommand{\INSTEB}{\affiliation{Institute for Nuclear Research of the Russian Academy of Sciences, Moscow, Russia}}
\newcommand{\INSTHA}{\affiliation{Kavli Institute for the Physics and Mathematics of the Universe (WPI), Todai Institutes for Advanced Study, University of Tokyo, Kashiwa, Chiba, Japan}}
\newcommand{\INSTCC}{\affiliation{Kobe University, Kobe, Japan}}
\newcommand{\INSTCD}{\affiliation{Kyoto University, Department of Physics, Kyoto, Japan}}
\newcommand{\INSTEJ}{\affiliation{Lancaster University, Physics Department, Lancaster, United Kingdom}}
\newcommand{\INSTFC}{\affiliation{University of Liverpool, Department of Physics, Liverpool, United Kingdom}}
\newcommand{\INSTFI}{\affiliation{Louisiana State University, Department of Physics and Astronomy, Baton Rouge, Louisiana, U.S.A.}}
\newcommand{\INSTJ}{\affiliation{Universit\'e de Lyon, Universit\'e Claude Bernard Lyon 1, IPN Lyon (IN2P3), Villeurbanne, France}}
\newcommand{\INSTHB}{\affiliation{Michigan State University, Department of Physics and Astronomy,  East Lansing, Michigan, U.S.A.}}
\newcommand{\INSTCE}{\affiliation{Miyagi University of Education, Department of Physics, Sendai, Japan}}
\newcommand{\INSTDF}{\affiliation{National Centre for Nuclear Research, Warsaw, Poland}}
\newcommand{\INSTFJ}{\affiliation{State University of New York at Stony Brook, Department of Physics and Astronomy, Stony Brook, New York, U.S.A.}}
\newcommand{\INSTGJ}{\affiliation{Okayama University, Department of Physics, Okayama, Japan}}
\newcommand{\INSTCF}{\affiliation{Osaka City University, Department of Physics, Osaka, Japan}}
\newcommand{\INSTGG}{\affiliation{Oxford University, Department of Physics, Oxford, United Kingdom}}
\newcommand{\INSTBB}{\affiliation{UPMC, Universit\'e Paris Diderot, CNRS/IN2P3, Laboratoire de Physique Nucl\'eaire et de Hautes Energies (LPNHE), Paris, France}}
\newcommand{\INSTGC}{\affiliation{University of Pittsburgh, Department of Physics and Astronomy, Pittsburgh, Pennsylvania, U.S.A.}}
\newcommand{\INSTFA}{\affiliation{Queen Mary University of London, School of Physics and Astronomy, London, United Kingdom}}
\newcommand{\INSTE}{\affiliation{University of Regina, Department of Physics, Regina, Saskatchewan, Canada}}
\newcommand{\INSTGD}{\affiliation{University of Rochester, Department of Physics and Astronomy, Rochester, New York, U.S.A.}}
\newcommand{\INSTBC}{\affiliation{RWTH Aachen University, III. Physikalisches Institut, Aachen, Germany}}
\newcommand{\INSTFB}{\affiliation{University of Sheffield, Department of Physics and Astronomy, Sheffield, United Kingdom}}
\newcommand{\INSTDI}{\affiliation{University of Silesia, Institute of Physics, Katowice, Poland}}
\newcommand{\INSTEH}{\affiliation{STFC, Rutherford Appleton Laboratory, Harwell Oxford,  and  Daresbury Laboratory, Warrington, United Kingdom}}
\newcommand{\INSTCH}{\affiliation{University of Tokyo, Department of Physics, Tokyo, Japan}}
\newcommand{\INSTBJ}{\affiliation{University of Tokyo, Institute for Cosmic Ray Research, Kamioka Observatory, Kamioka, Japan}}
\newcommand{\INSTCG}{\affiliation{University of Tokyo, Institute for Cosmic Ray Research, Research Center for Cosmic Neutrinos, Kashiwa, Japan}}
\newcommand{\INSTGI}{\affiliation{Tokyo Metropolitan University, Department of Physics, Tokyo, Japan}}
\newcommand{\INSTF}{\affiliation{University of Toronto, Department of Physics, Toronto, Ontario, Canada}}
\newcommand{\INSTB}{\affiliation{TRIUMF, Vancouver, British Columbia, Canada}}
\newcommand{\INSTG}{\affiliation{University of Victoria, Department of Physics and Astronomy, Victoria, British Columbia, Canada}}
\newcommand{\INSTDJ}{\affiliation{University of Warsaw, Faculty of Physics, Warsaw, Poland}}
\newcommand{\INSTDH}{\affiliation{Warsaw University of Technology, Institute of Radioelectronics, Warsaw, Poland}}
\newcommand{\INSTFD}{\affiliation{University of Warwick, Department of Physics, Coventry, United Kingdom}}
\newcommand{\INSTGE}{\affiliation{University of Washington, Department of Physics, Seattle, Washington, U.S.A.}}
\newcommand{\INSTGH}{\affiliation{University of Winnipeg, Department of Physics, Winnipeg, Manitoba, Canada}}
\newcommand{\INSTEA}{\affiliation{Wroclaw University, Faculty of Physics and Astronomy, Wroclaw, Poland}}
\newcommand{\INSTH}{\affiliation{York University, Department of Physics and Astronomy, Toronto, Ontario, Canada}}

\INSTC
\INSTEE
\INSTFE
\INSTD
\INSTGA
\INSTI
\INSTGB
\INSTFG
\INSTFH
\INSTBA
\INSTEF
\INSTEG
\INSTDG
\INSTCB
\INSTED
\INSTEC
\INSTEI
\INSTGF
\INSTBE
\INSTBF
\INSTBD
\INSTEB
\INSTHA
\INSTCC
\INSTCD
\INSTEJ
\INSTFC
\INSTFI
\INSTJ
\INSTHB
\INSTCE
\INSTDF
\INSTFJ
\INSTGJ
\INSTCF
\INSTGG
\INSTBB
\INSTGC
\INSTFA
\INSTE
\INSTGD
\INSTBC
\INSTFB
\INSTDI
\INSTEH
\INSTCH
\INSTBJ
\INSTCG
\INSTGI
\INSTF
\INSTB
\INSTG
\INSTDJ
\INSTDH
\INSTFD
\INSTGE
\INSTGH
\INSTEA
\INSTH

\author{K.\,Abe}\INSTBJ
\author{J.\,Adam}\INSTFJ
\author{H.\,Aihara}\INSTCH\INSTHA
\author{C.\,Andreopoulos}\INSTEH\INSTFC
\author{S.\,Aoki}\INSTCC
\author{A.\,Ariga}\INSTEE
\author{S.\,Assylbekov}\INSTFG
\author{D.\,Autiero}\INSTJ
\author{M.\,Barbi}\INSTE
\author{G.J.\,Barker}\INSTFD
% ...

\collaboration{The T2K Collaboration}\noaffiliation

\date{\today}% It is always \today, today,

\begin{abstract}
We report a measurement of the $\nu_\mu$ charged current quasi-elastic cross-sections on carbon in the T2K on-axis neutrino beam.
The measured charged current quasi-elastic cross-sections on carbon at mean neutrino energies of 1.94~GeV and 0.93~GeV are
$(11.95\pm 0.19(stat.)_{-1.47}^{+1.82}
(syst.))\times 10^{-39}\mathrm{cm}^2/\mathrm{neutron}$,
and $(10.64\pm 0.37(stat.)_{-1.65}^{+2.03}
(syst.))\times 10^{-39}\mathrm{cm}^2/\mathrm{neutron}$, respectively. These results agree well with the predictions of neutrino interaction models.
In addition, we investigated the effects of the nuclear model and the multi-nucleon interaction.

%\begin{description}
%\end{description}
\end{abstract}

\pacs{Valid PACS appear here}% PACS, the Physics and Astronomy
\maketitle

%\tableofcontents

\section{Introduction}\label{sec:introduction}
The T2K (Tokai-to-Kamioka) experiment is a long baseline neutrino oscillation experiment \cite{t2k_nim} whose primary goal is
a precise measurement of the neutrino oscillation parameters via the appearance of electron neutrinos and the disappearance of muon neutrinos \cite{t2k_joint_oa}.
An almost pure intense muon neutrino beam is produced at J-PARC (Japan Proton Accelerator Research Complex) in Tokai.
The proton beam impinges on a graphite target to produce charged pions, which are focused by three magnetic horns \cite{horn_ichikawa}. The pions decay mainly into muon--muon-neutrino pairs during their passage through the 96-meter decay volume.
The neutrinos are measured by the near detectors (INGRID \cite{ingrid_nim} and ND280 \cite{p0d_nim,tpc_abgrall,fgd_amaudruz,ecal_jinst,smrd_nim}) on the J-PARC site and the far detector (Super-Kamiokande \cite{sk_detector}) in Kamioka, located 295\,km away from J-PARC.

A precise neutrino oscillation measurement requires good knowledge of neutrino interaction cross-sections.
The neutrino charged current quasi-elastic (CCQE) scattering is especially important for T2K because it is used as the signal mode for the T2K neutrino oscillation measurement.
The $\nu_\mu$ CCQE cross-section on carbon was measured by MiniBooNE \cite{miniboone_ccqe}, SciBooNE \cite{sciboone_ccqe}, NOMAD \cite{nomad_ccqe}, MINER$\nu$A \cite{minerva_ccqe_nu} and LSND \cite{lsnd_ccqe}.
The Llewellyn Smith formalism \cite{smith} using the relativistic Fermi gas model \cite{moniz} is generally used to describe CCQE scattering with a neutron in the nucleus.
However, this approach does not provide a good description of 
existing data.  Several modifications to the model have been 
proposed to account for the discrepancies, but none of them has yet achieved general acceptance.
One of the more promising approaches involves the introduction of neutrino interactions with two or more nucleons via so-called meson exchange current (MEC) into the neutrino interaction model \cite{mec_martini1, mec_martini2, mec_nieves1, mec_nieves2, mec_bodek, mec_lalakulich, mec_amaro}.
On the other hand, uncertainties in the nuclear model are also regarded as a possible cause of the discrepancy \cite{ankowski}.
In order to resolve the puzzle, additional CCQE cross-section measurements are required.

In this paper, we present a measurement of the $\nu_\mu$ CCQE cross-section on carbon at neutrino energies around 1~GeV using the INGRID detector.
The CCQE signal is defined as the conventional two-body interaction with a single nucleon.
We selected the CCQE candidate events in INGRID and estimated the CCQE cross-section by subtracting background and correcting for selection efficiency based on the NEUT neutrino interaction generator \cite{neut_hayato}.
The CCQE cross-section was estimated assuming two different nuclear models, and with and without the multi-nucleon interaction in order to check their effects on the cross-section result.

T2K collected data corresponding to $7.32\times10^{20}$ protons on target (POT) during the five run periods listed in Table~\ref{dataset}.
For this cross-section measurement, data from Run 2, 3c and 4 are used.
The total data set for the cross-section measurement corresponds to $6.04\times10^{20}$ POT.

\begin{table}[htbp]
\begin{center}
  \caption{T2K data-taking periods and integrated
protons on target (POT). Data of Run 1, 3b, 5a, 5b were not used for the cross-section measurement.}
  \begin{tabular}{cccc}
    \hline\hline
     Run & Dates & Horn & Integrated\\ 
     period & & current & POT \\ 
    \hline
     (Run 1) & Jan. 2010 $-$ Jun. 2010 & 250kA &$0.32\times10^{20}$\\ 
     Run 2 & Nov. 2010 $-$ Mar. 2011 & 250kA &$1.11\times10^{20}$\\
     (Run 3b) & Mar. 2012 & 205kA &$0.21\times10^{20}$\\
     Run 3c & Apr. 2012 $-$ Jun. 2012 & 250kA &$1.37\times10^{20}$\\
     Run 4 & Oct. 2012 $-$ May. 2013 & 250kA &$3.56\times10^{20}$\\
     (Run 5a) & May. 2014 $-$ Jun. 2014 & 250kA & $0.24\times10^{20}$\\
     (Run 5b) & Jun. 2014 & $-$250kA & $0.51\times10^{20}$\\
    \hline\hline
  \end{tabular}
  \label{dataset}
  \end{center}
\end{table}

The remainder of this paper is organized as follows:
Details of the INGRID detector and Monte Carlo simulations are explained in Sec.\ \ref{sec:detector} and \ref{sec:mc}, respectively.
Section \ref{sec:event_selection} summarizes the CCQE event selection.
The analysis method of the cross-section measurement is described in Sec.\ \ref{sec:method}.
Section \ref{sec:error} describes the systematic errors.
The results and conclusions are given in Sec.\ \ref{sec:result} and \ref{sec:conclusion}, respectively.

\section{Detector configuration}\label{sec:detector}
The INGRID (Interactive Neutrino GRID) detector is an on-axis neutrino detector located 280\,m downstream of the proton target while ND280 and Super-Kamiokande are located 2.5$^\circ$ off the beamline axis.
It consists of 16 identical standard modules and an extra module called the Proton Module.

The main purpose of the standard modules is to monitor the neutrino beam direction.
Each of the modules consists of nine iron target plates and eleven tracking scintillator planes.

\begin{figure}[htbp]
  \begin{center}
  \includegraphics[width=68mm]{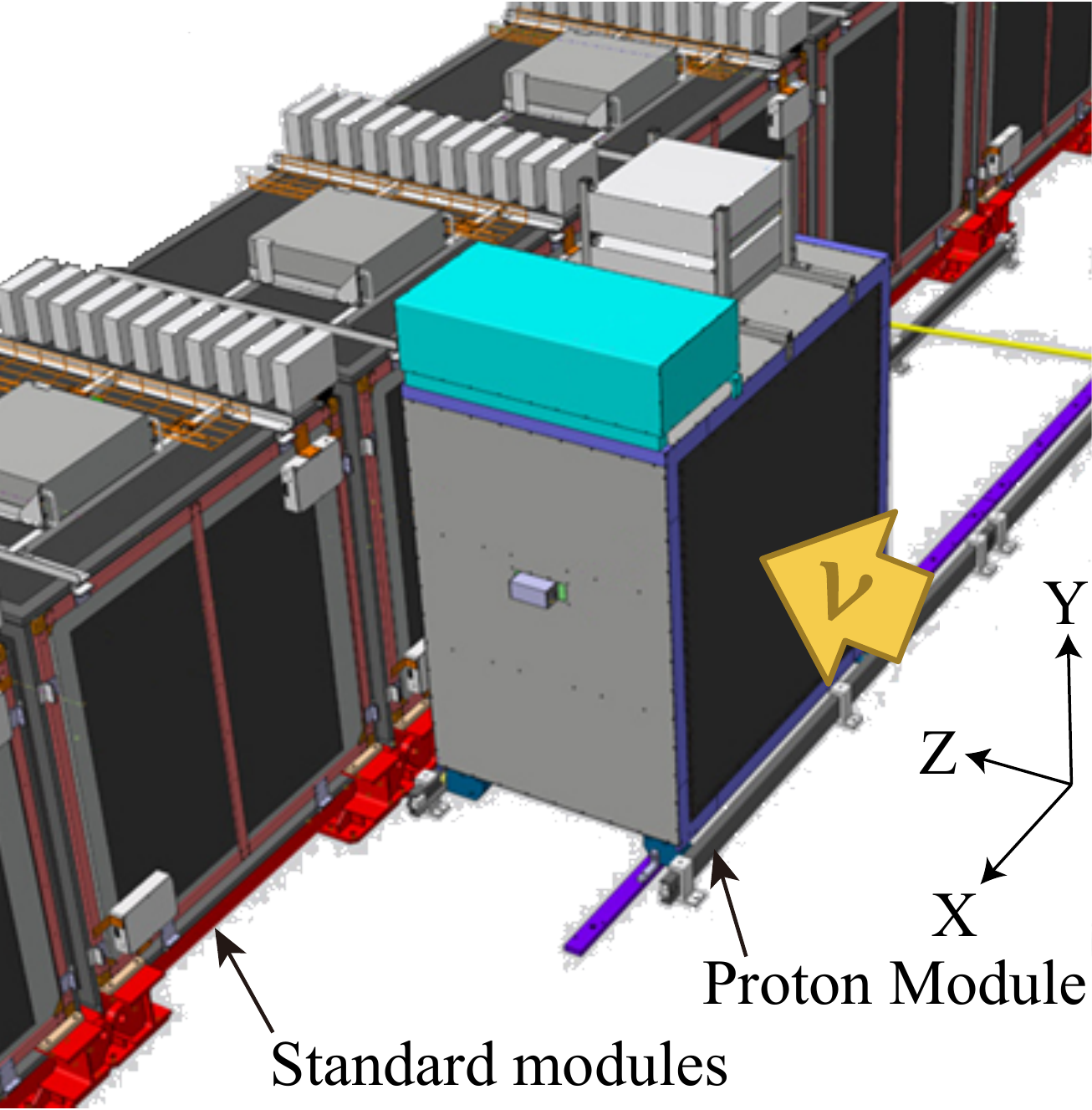}
    \caption{Schematic view of the Proton Module and the standard modules.}
  \label{pm_ingrid}
  \end{center}
\end{figure}

\begin{figure}[htbp]
	\begin{center}
		\includegraphics[width=72mm]{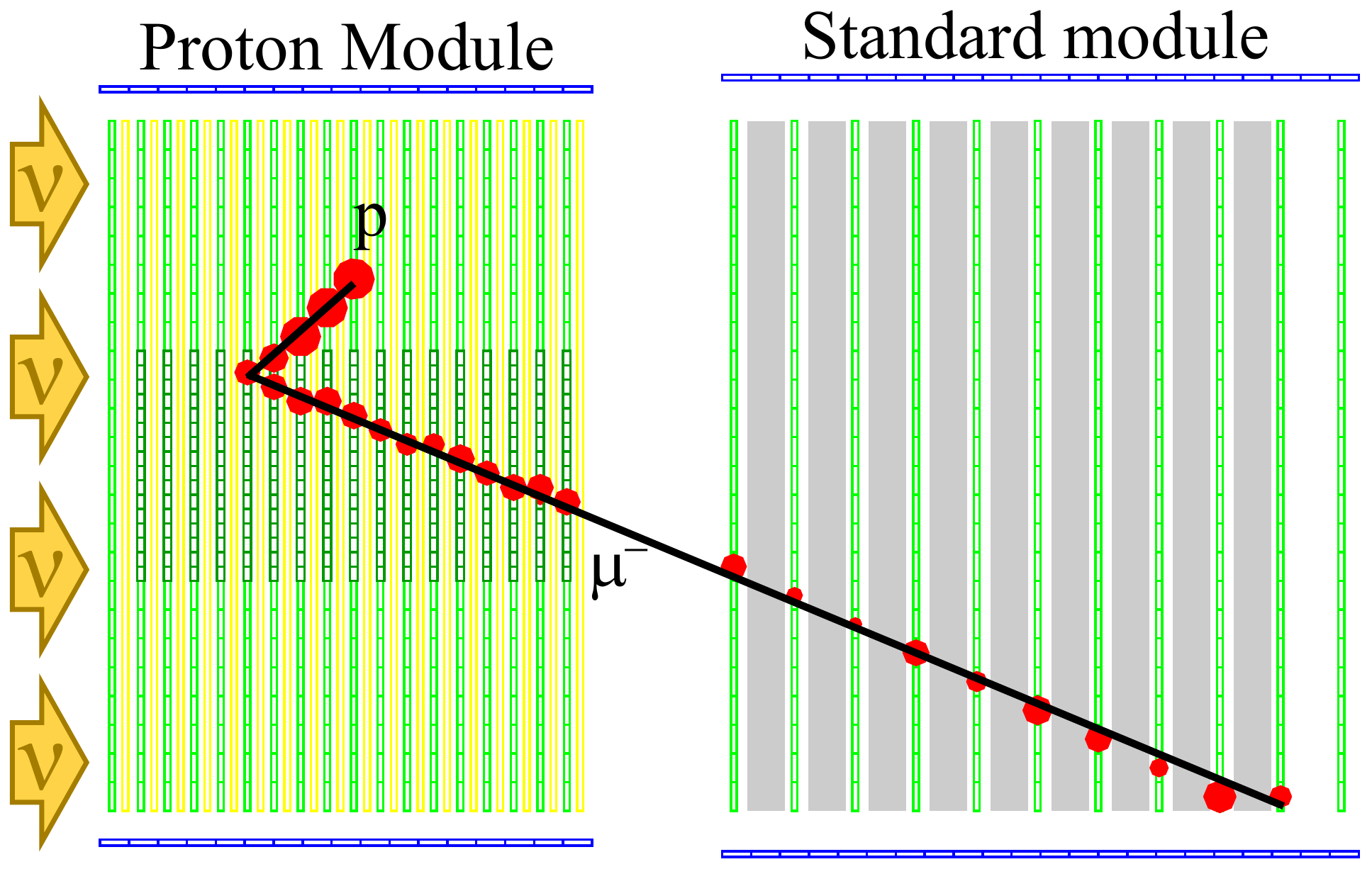}
		\caption{Event display of an MC CCQE event in the Proton Module.}
		\label{display_ccqe}
	\end{center}
\end{figure}

In contrast, the Proton Module was developed specifically for the neutrino cross-section measurement.
It is a fully-active tracking detector which consists of 36 tracking layers surrounded by veto planes to reject charged particles coming from outside of the modules.
The tracking layers also serve as the neutrino interaction target.
The total target mass in the fiducial volume is 303~kg.
Seven of the 16 standard modules are horizontally aligned, and the Proton Module is placed in front of the central module of them.
In this cross-section measurement, the Proton Module is used as the neutrino interaction target, and the seven standard modules located downstream of the Proton Module are used as the muon detector.
The schematic view of the Proton Module and the standard modules, and an event display of an MC CCQE event are shown in Figs.\ \ref{pm_ingrid} and \ref{display_ccqe}, respectively.

The standard module and the Proton Module use a common readout system.
Scintillation light is collected
and transported to a photodetector with a wavelength shifting fiber (WLS fiber) which is inserted in a hole at the center of the scintillator strip.
The light is read out by a Multi-Pixel Photon Counter (MPPC)
\cite{mppc_yokoyama1, mppc_yokoyama2} attached to one end of the WLS fiber.
The integrated charge and timing information from each MPPC is digitized by the Trip-t front-end board (TFB) \cite{tfb}.
The integration cycle is synchronized with the neutrino beam pulse structure.
Details of the components and the basic performance of the INGRID detector are described in Ref.\ \cite{ingrid_nim}.

\section{Monte Carlo simulation}\label{sec:mc}
The INGRID Monte Carlo (MC) simulation consists of three main parts: a simulation of the neutrino beam
production, which predicts the neutrino flux and energy spectrum of each neutrino flavor; a neutrino interaction simulation, which is used to calculate the neutrino interaction cross-sections and the kinematics of the final state particles taking into account the intranuclear interactions of hadrons;
and a detector response simulation which reproduces the final-state particles' motion and interaction with material, the scintillator light yield, and the response of the WLS fibers, MPPCs, and front-end electronics.

\subsection{Neutrino beam prediction}
\begin{figure}[htbp]
  \begin{center}
  \includegraphics[width=71mm]{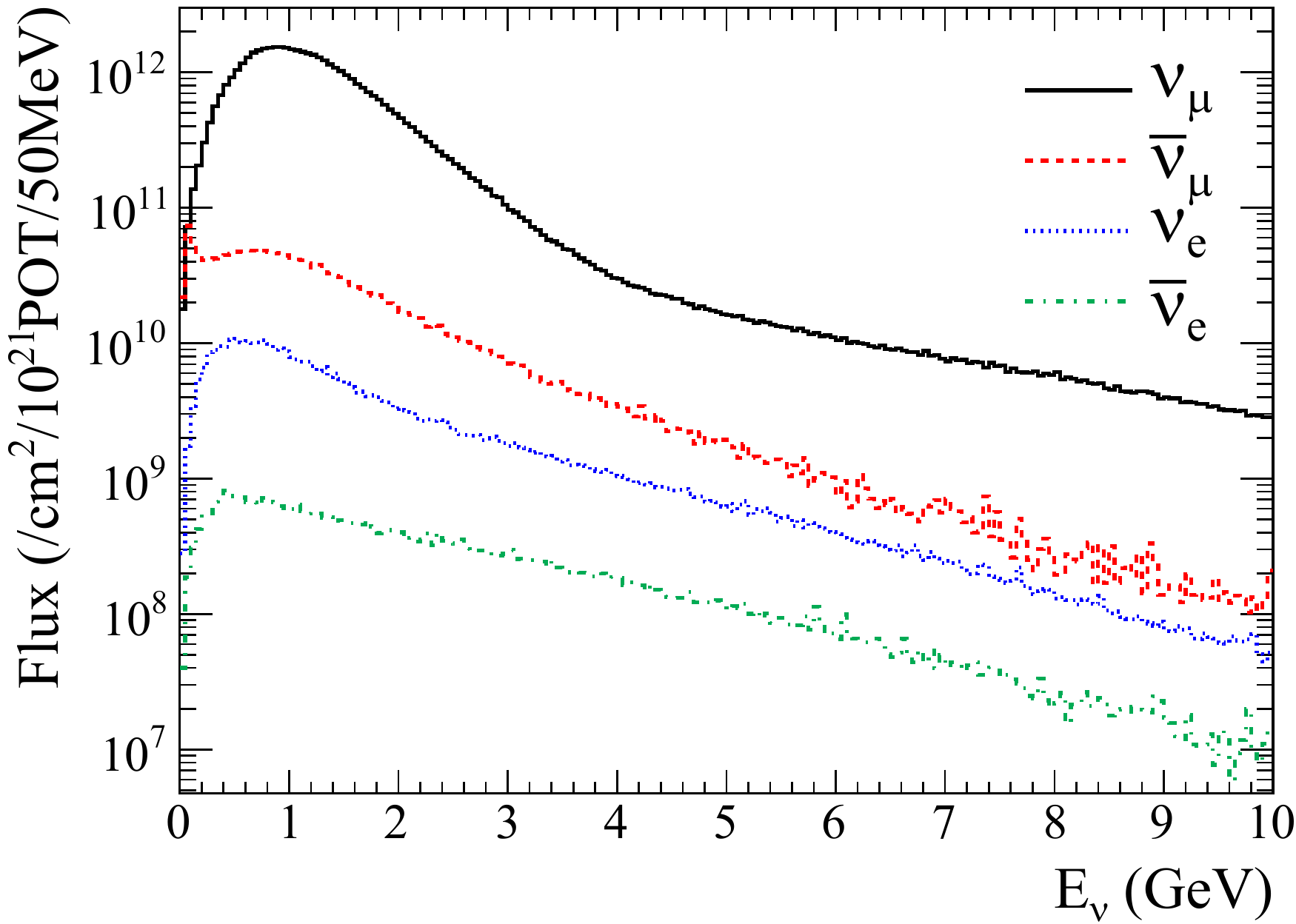}
  \caption{Neutrino energy spectrum for each neutrino species at the Proton Module with the 250kA horn current predicted by JNUBEAM.}
  \label{flux_flavor}
  \end{center}
\end{figure}

To predict the neutrino fluxes and energy spectra, a neutrino beam Monte Carlo simulation, called JNUBEAM \cite{flux_prediction}, was developed based on the GEANT3 framework \cite{geant3}. We compute the neutrino beam fluxes starting from models (FLUKA2008 \cite{fluka1,fluka2} and GCALOR \cite{gcalor}) and tune them using existing hadron production data (NA61/SHINE \cite{na61_1, na61_2}, Eichten {\it et al.} \cite{eichten} and Allaby {\it et al.} \cite{allaby}).
Since we use only the beam data with the 250kA horn current, the horn current in the simulation is fixed at 250kA.
The predicted neutrino energy spectra at the center of INGRID are shown in Fig.\ \ref{flux_flavor}. Energy spectra 10\,m upstream of INGRID are predicted with the same procedure in order to simulate the background events from neutrino interactions in the walls of the experimental hall.

\subsection{Neutrino interaction simulation}
\begin{figure}[htbp]
  \begin{center}
  \includegraphics[width=53mm, angle=90]{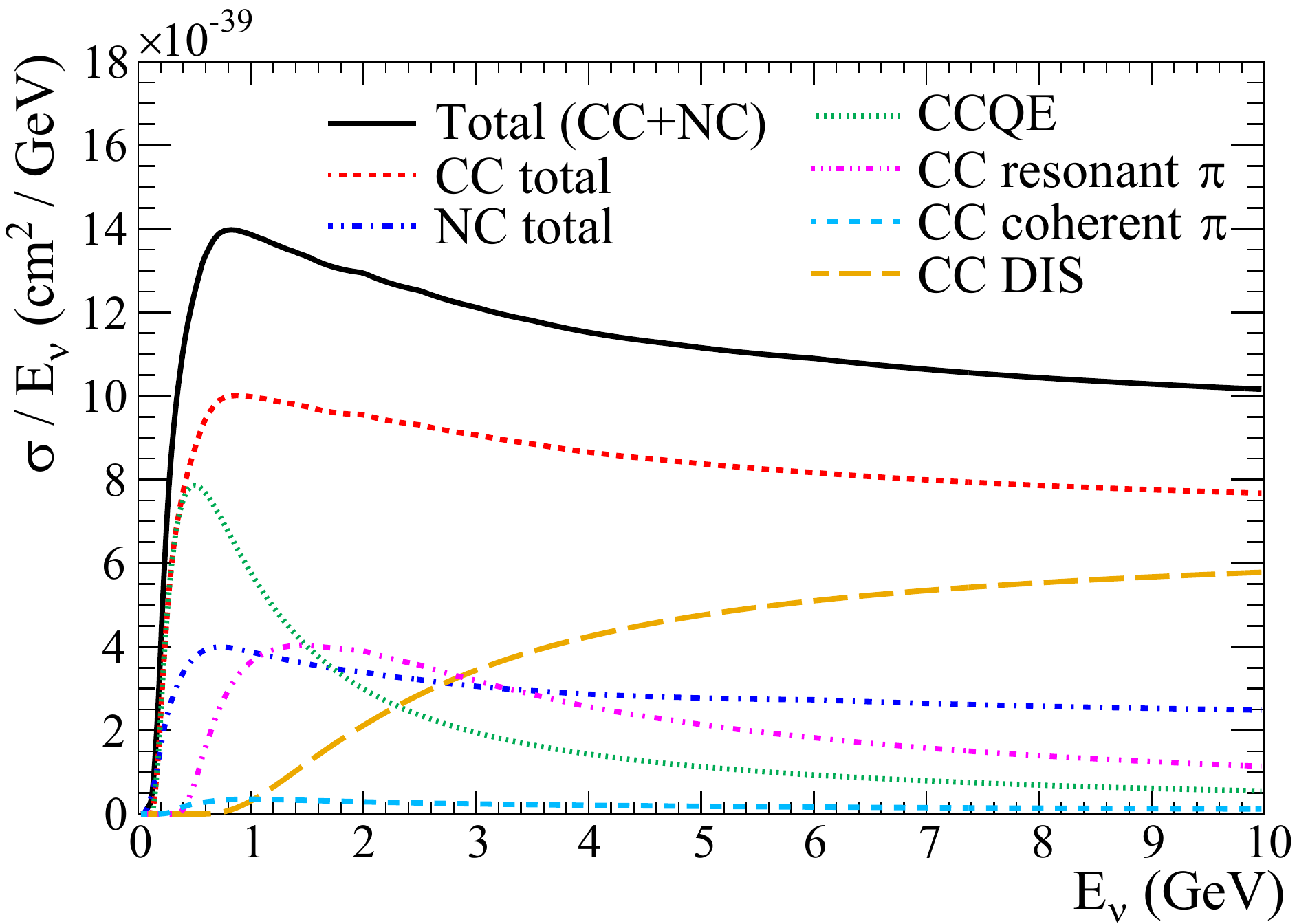}
  \caption{Neutrino-nucleus cross-sections per nucleon of carbon nucleus divided by the neutrino energy, as predicted by NEUT.}
  \label{xsec_neut}
  \end{center}
\end{figure}

Neutrino interactions with nuclear targets are simulated with the NEUT program library \cite{neut_hayato}.
Both the primary neutrino interactions in nuclei and the secondary interactions of the hadrons in the nuclear medium are simulated in NEUT\@.
NEUT uses the Llewellyn Smith formalism \cite{smith} with the relativistic Fermi gas model for quasi-elastic scattering, the Rein-Sehgal model \cite{rands_res, rands_coh} for resonant meson production and coherent $\pi$ production and GRV98 (Gl\"{u}ck-Reya-Vogt-1998) \cite{gluck} parton distributions with Bodek-Yang modifications \cite{bodek, yang} for deep inelastic scattering.
For the measurement presented in this paper, the final state interactions of the nucleons in nuclei are important, because we use information about protons produced in the interaction.
In NEUT, both elastic scattering and pion production are considered.
The differential cross-sections were obtained from nucleon-nucleon scattering experiments~\cite{bertini}. For pion production, the isobaric nucleon model~\cite{lindenbaum} is used.
The $\pi$-less $\Delta$ decay, which is the interaction of the $\Delta$ within the target nucleus prior to decay into a pion and a nucleon, is also an important process because the resonant interaction event is often misidentified as the CCQE event due to this interaction. We use the data from Refs.~\cite{pdd_1, pdd_2} to determine the probability of the number of emitted nucleons and the kinematics of the two-nucleon emission in the $\pi$-less $\Delta$ decay. For emission of three or more nucleons, the nucleons are isotropically emitted.
The simulations in NEUT are described in more detail in Refs.\ \cite{t2k_ccinc, neut_hayato}.
Figure~\ref{xsec_neut} shows the neutrino-nucleus cross-sections per nucleon divided by the neutrino energy predicted by NEUT\@.
Additionally, a CCQE cross-section prediction by a different neutrino interaction simulation package, GENIE \cite{genie}, is used for comparison.
GENIE also uses the Llewellyn Smith formalism with the relativistic Fermi gas model.
However, the nominal value of the axial mass \cite{m_a_ref} differs from that in NEUT (1.21~GeV/$c^2$ for NEUT and 0.99~GeV/$c^2$ for GENIE).
In addition, GENIE incorporates short range nucleon-nucleon correlations in the relativistic Fermi gas model and handles kinematics for off-shell scattering according to the model of Bodek and Ritchie \cite{bandr} while NEUT uses the Smith-Moniz model \cite{moniz}.

\subsection{INGRID detector response simulation}
The INGRID detector simulation was developed using the Geant4 framework \cite{geant4}.
It models the real detector structures (geometries, materials).
The structure of the walls of the experimental hall is also modeled to simulate background events from neutrino interactions in the walls.
The particles' motion and interactions with the materials are simulated, and the energy deposit of each particle inside the scintillator is stored.
Simulations of hadronic interactions are performed with the QGSP BERT physics list \cite{qgsp_bert}.
The energy deposit is converted into a number of photons.
Quenching effects of the scintillation are modeled based on Birks' law \cite{birk1,birk2}.
The effect of collection and attenuation of the light in the scintillator and the WLS fiber is modeled based on the results of electron beam irradiation tests.
The non-linearity of the MPPC response is also taken into account, since the number of detectable photoelectrons is limited by the number of MPPC pixels. The number of photoelectrons is smeared according to statistical fluctuations and electrical noise.
The dark count of the MPPCs is added with a probability calculated from the measured dark rate.
Because the response of the ADCs on front-end electronics is not linear, its response is modeled based on the results of a charge injection test.

\section{Event selection}\label{sec:event_selection}
\subsection{CC event selection}
As the first step, CC interaction events in the Proton Module are reconstructed and selected as follows where the coordinates shown in Fig. \ref{pm_ingrid} are used.
\begin{enumerate}
 \item Two-dimensional XZ and YZ tracks in the Proton Module and the standard modules are reconstructed independently from hit information in each plane using a specially developed reconstruction algorithm.
 \item Track matching between the Proton Module tracks and the standard module tracks are performed.
 \item Three-dimensional tracks are searched for among pairs of two-dimensional XZ tracks and YZ tracks by matching the Z positions of the track edges.
 \item The neutrino interaction vertices are searched for by looking at the upstream edges of the three-dimensional tracks.
 \item Events within $\pm$100~nsec from the expected timing, which is calculated from the timing of the pulsed primary proton beam, are selected.
 \item Events which have a hit in a veto plane at the upstream position extrapolated from a reconstructed track are rejected. The events rejected by the front veto plane are identified as beam induced muon backgrounds. The number of neutrino interactions on the walls in the MC simulation is normalized by the observed number of the beam induced muon backgrounds.
 \item Events having a vertex outside the fiducial volume, which is defined as within $\pm$50~cm from the module center in the X and Y directions, are rejected.
\end{enumerate}
Further details of this neutrino event selection is written in Ref.\cite{ingrid_ccincl}.
After this selection, CCQE events make up 37.67\% of the MC sample.
To increase the selection purity for CCQE events, additional cuts are applied based on the number of reconstructed tracks, the $dE/dx$ particle identification variable and the reconstructed event kinematics.
Then the selected CCQE candidate events are classified according to the neutrino energy.

\subsection{Number of tracks}
\subsubsection{Number of tracks from the vertex}
\begin{figure}[htbp]
 \begin{center}
    \includegraphics[width=76mm]{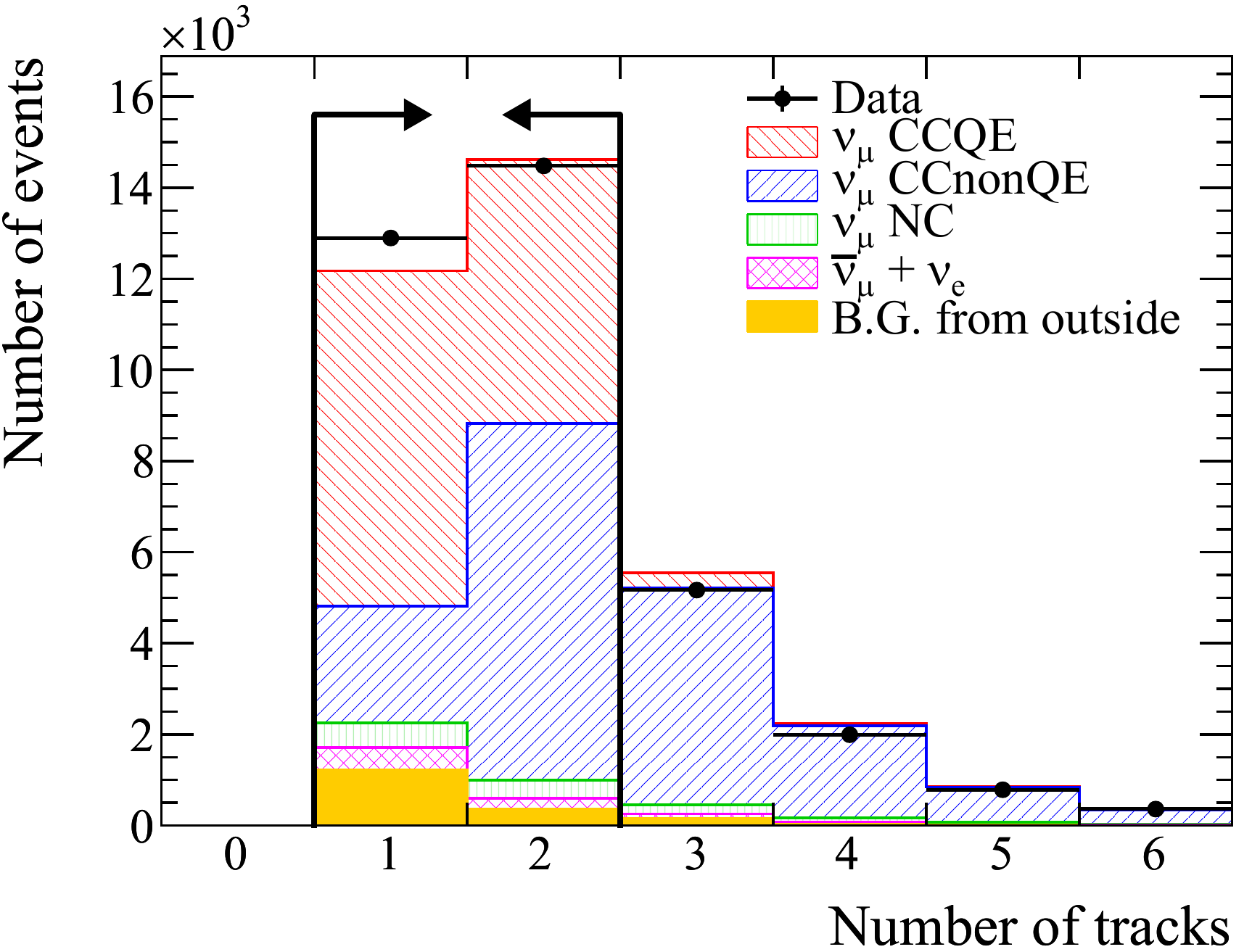}
  \caption{The number of reconstructed tracks from vertex in the Proton Module. The colored histograms are the MC predictions divided by the neutrino type and interaction type. The neutrino interaction events in the Proton Module are absolutely normalized to POT, and the background events from outside are normalized to beam induced muon backgrounds.}
\label{ntrack_ccqe}
   \end{center}
\end{figure}

The CCQE interaction produces two particles inside the target nucleus, a muon and a proton.  However, the proton undergoes final state interactions (FSI) in the residual nucleus and does not always escape unaltered.  In addition, the proton may not be reconstructed in the detector due to its short range.
Thus, events with either one or two reconstructed tracks coming from the vertex are selected (Fig.~\ref{ntrack_ccqe}).
Hereafter, the events with one and two reconstructed tracks are referred to as the one-track sample and the two-track sample, respectively.
The analysis must account for the above effects to produce the final cross-section result from each sample.

\subsubsection{Number of matched tracks}
\begin{figure}[htbp]
 \begin{center}
    \includegraphics[width=76mm]{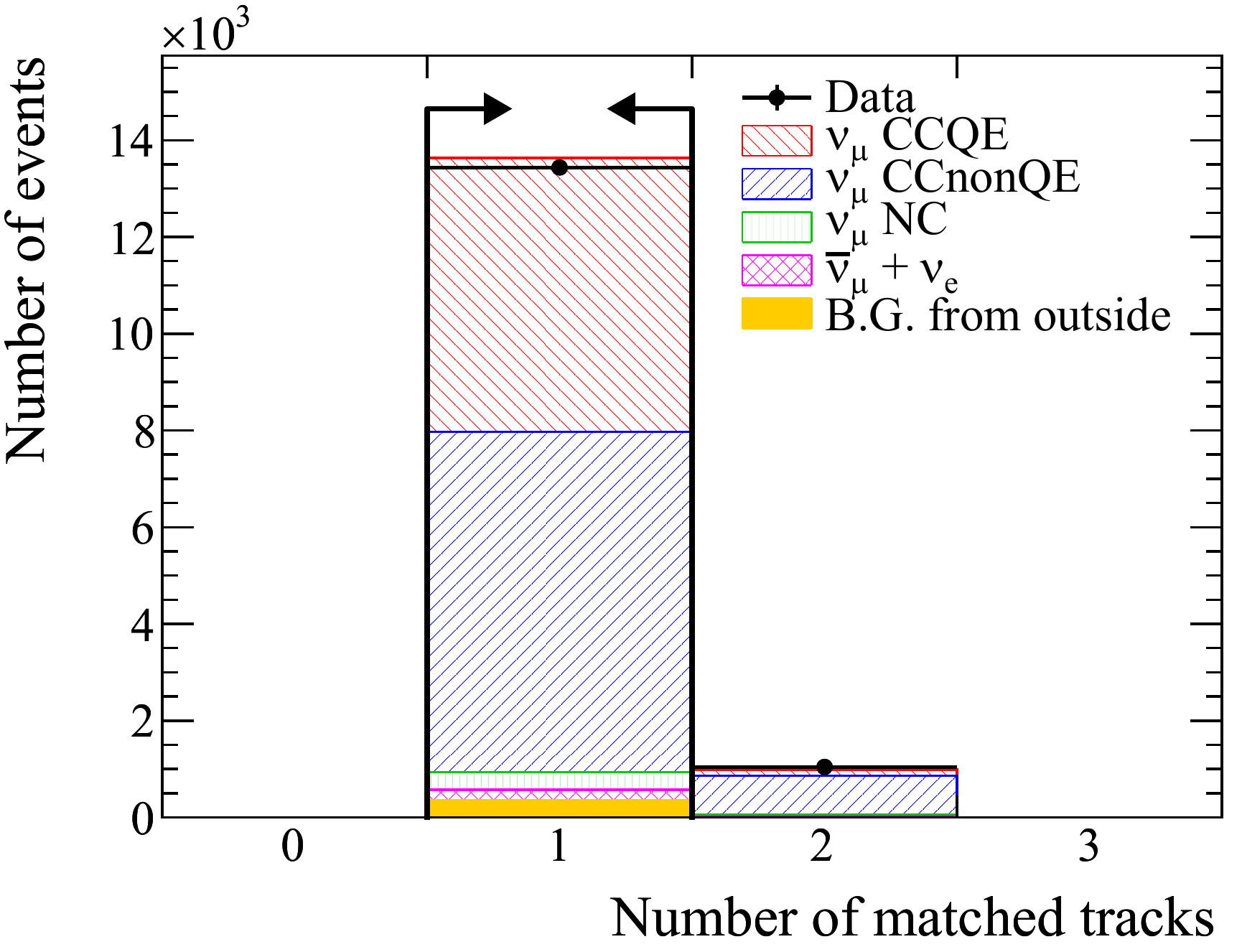}
  \caption{The number of matched tracks between the Proton Module and the standard module for the two track sample.}
\label{ningtrack_ccqe}
   \end{center}
\end{figure}

When selecting CC events, events with at least one
matched track between the Proton Module and the standard module are accepted in order to select those containing a long muon track.
About 7\% of the selected events in the two-track sample have two matched tracks.
Simulations indicate that the second matched track is usually a pion from a CC-nonQE interaction such as CC resonant pion production, CC coherent pion production etc.
Thus, events with exactly one matched track are selected for the two-track sample (Fig.\ \ref{ningtrack_ccqe}).
Hereafter, the matched track is referred to as the first track and the remaining track as the second track.

\subsection{Particle identification}
\subsubsection{Definition of the muon confidence level (MuCL) variable}
Particle identification (PID) based on $dE/dx$ information is applied on both the one-track sample and the two-track sample.
The $dE/dx$ for each scintillator plane is calculated from the light yield divided by the path length of the track in the scintillator where the light attenuation in the WLS fiber is corrected.
The first step of the particle identification is to estimate a confidence level that a particle is a muon on a plane-by-plane basis.
The confidence level at each plane is defined as the fraction of events in the expected $dE/dx$ distribution of muons above the observed $dE/dx$ value because muons have a lower $dE/dx$ value than protons.
The expected $dE/dx$ distribution of muons is obtained from the beam-induced muon backgrounds which are mainly created by the neutrino interactions in the walls of the detector hall.
The cumulative distribution function of the muon $dE/dx$ distribution corresponds to the confidence level.
The calculated confidence level at the $i$-th plane as a function of $dE/dx$ is referred to as $CL_i$.

The next step is to combine the confidence levels ($CL_i$) obtained from all the planes penetrated by the track to form a total confidence level.
In the case where the track penetrates only two planes, the procedure to combine the two confidence levels, $CL_1$ and $CL_2$, is as follows.
Assuming the confidence levels to be independent of one another, the combined probability is the product, $P=CL_1\times CL_2$.
In the $xy$-plane of the two confidence levels, the hyperbola $xy=P$ gives such a combined probability and the unified muon confidence level, MuCL, is the fraction of possible $x, y$ values that give $xy<P$.
MuCL is expressed as
\begin{eqnarray}
\mathrm{MuCL}&=&1-\int_{P}^{1}dx\int_{P/x}^{1}dy \nonumber \\
&=&P(1-\mathrm{ln}P).
\end{eqnarray}
In analogy with the two-plane case, the muon confidence level combined from $n$ planes is expressed as
\begin{equation}
\mathrm{MuCL} = P \times \sum_{i=0}^{n-1}\frac{(-\ln P)^i}{i!},\; P = \prod_{i=1}^{n}CL_i.
\end{equation}
When a Proton Module track is matched with a standard module track, the standard module track is also used to make the MuCL.

\subsubsection{Separation of muon-like tracks and proton-like tracks}
\begin{figure}[htbp]
 \begin{center}
    \includegraphics[width=76mm]{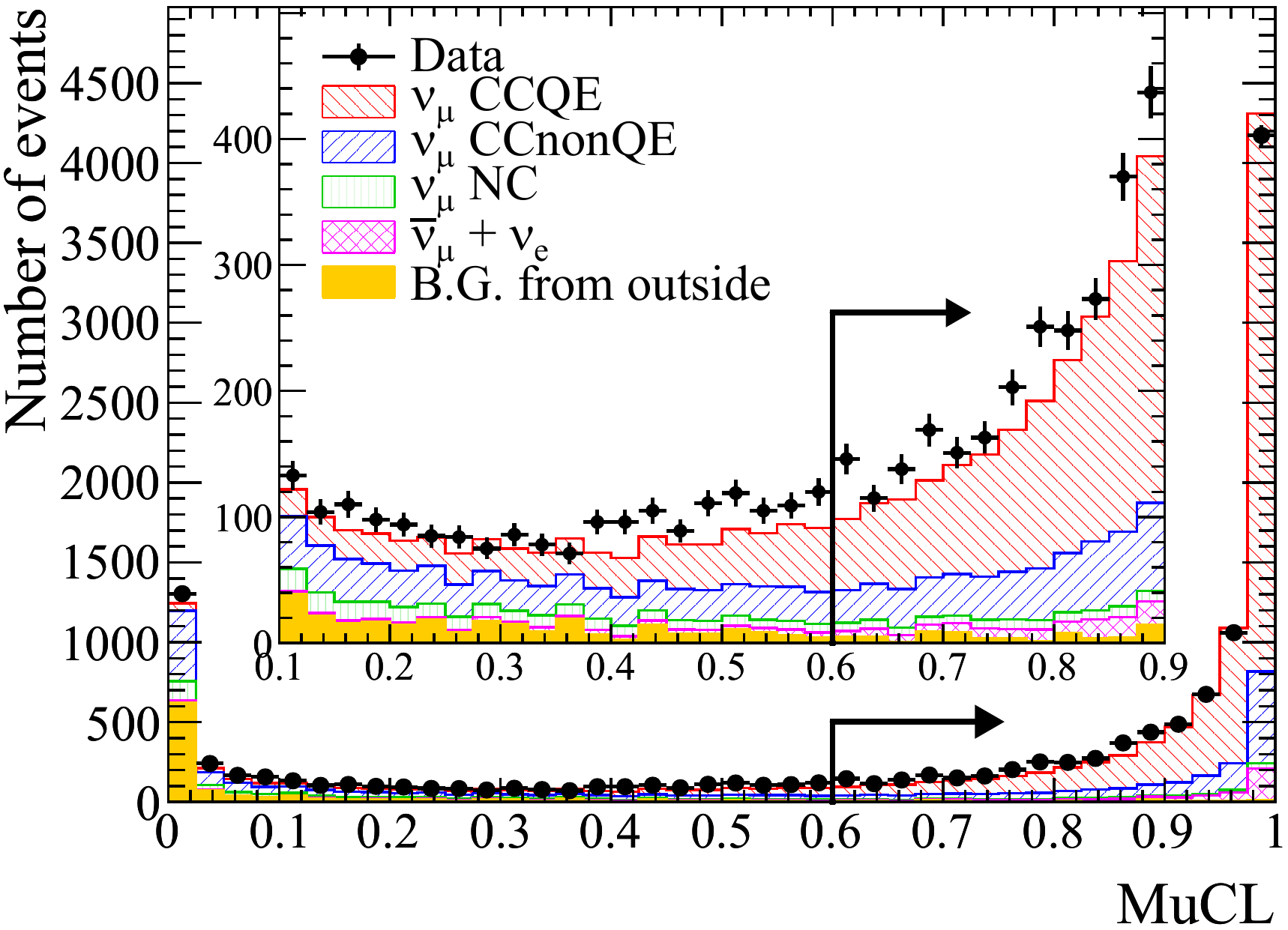}
  \caption{MuCL distributions for the one-track sample.}
\label{mumucl1}
   \end{center}
\end{figure}

\begin{figure}[htbp]
 \begin{center}
    \includegraphics[width=76mm]{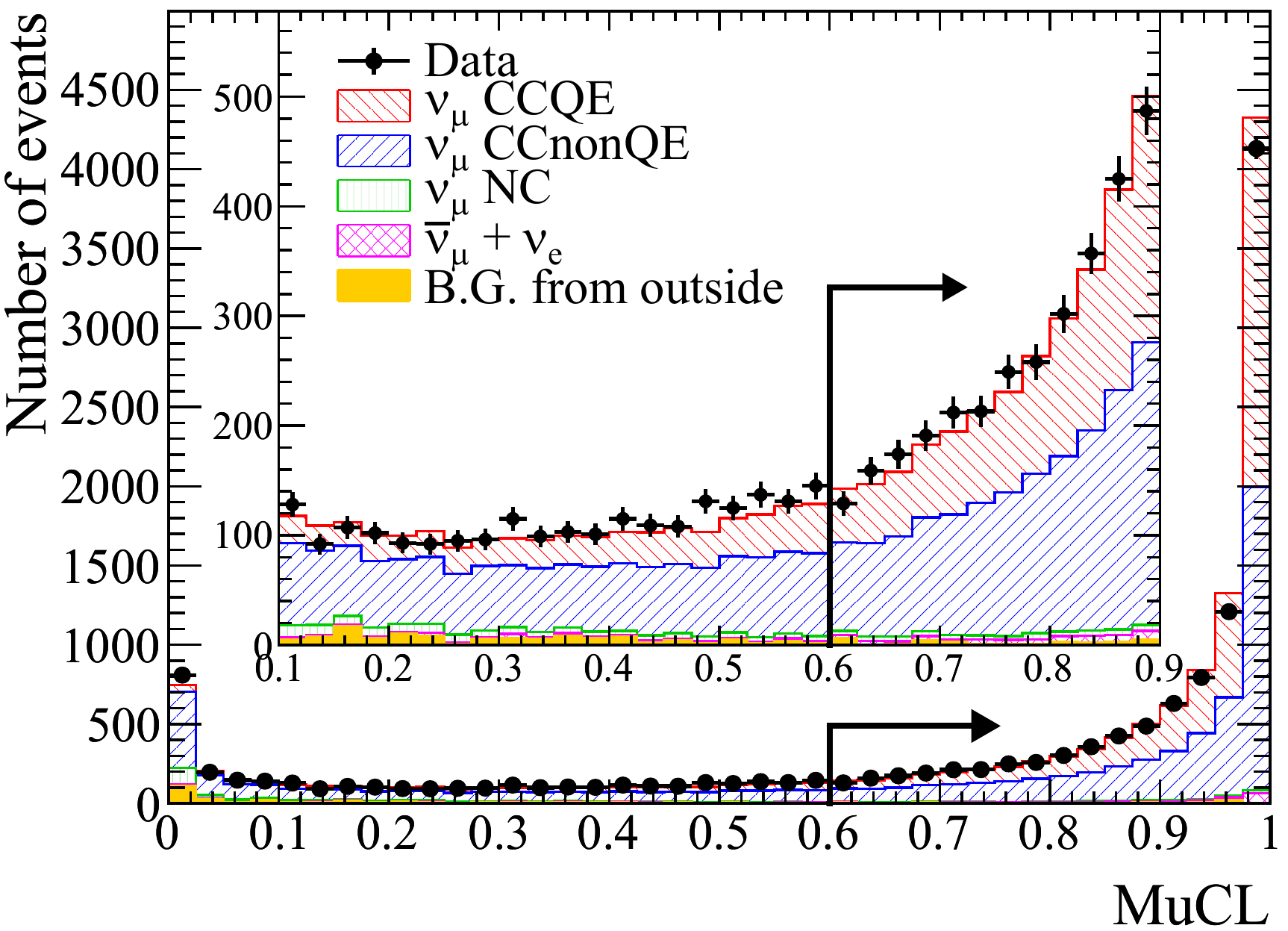}
  \caption{MuCL distributions of the first track for the two-track sample.}
\label{mumucl2}
   \end{center}
\end{figure}

\begin{figure}[htbp]
 \begin{center}
    \includegraphics[width=76mm]{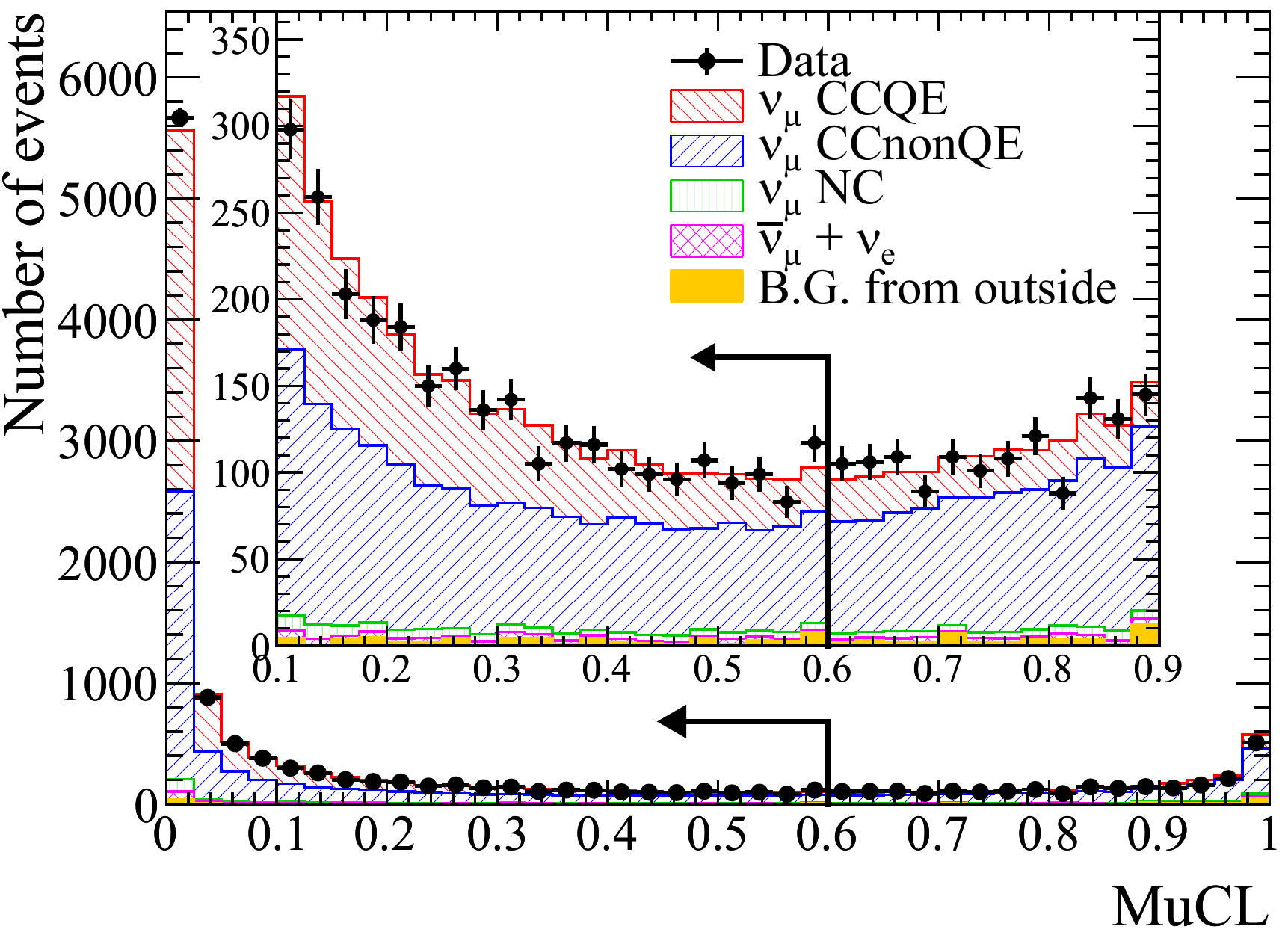}
  \caption{MuCL distributions of the second track for the two-track sample.}
\label{pmucl2}
   \end{center}
\end{figure}

The last step of the particle identification is to distinguish the tracks using the MuCL.
In this analysis, tracks whose MuCL are more than 0.6 are identified as muon-like and those less than 0.6 are identified as proton-like.
The probability of misidentifying a muon track (a proton track) as proton-like (muon-like) in the MC simulation is 12.5\% (10.9\%).
Most pion tracks are identified as muon-like, since the mass of the pion is similar to the muon mass.
For the one-track sample, events having a muon-like track (Fig.~\ref{mumucl1}) are selected as the CCQE enhanced sample.
For the two-track sample, the first track is required to be muon-like and the second track to be proton-like (Fig.~\ref{mumucl2}, \ref{pmucl2}).

\subsection{Kinematic cut}
\begin{figure}[htbp]
  \begin{center}
  \includegraphics[width=80mm]{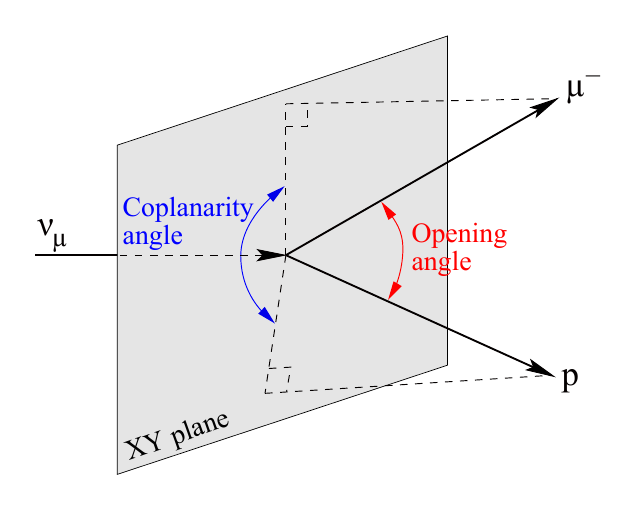}
  \caption{Definition of the coplanarity angle and the opening angle.}
  \label{kinematic}
  \end{center}
\end{figure}

In addition, two kinematic cuts are applied to the two-track sample.
These cuts use two angles called the coplanarity angle and the opening angle, defined as shown in Fig.~\ref{kinematic}.

\subsubsection{Coplanarity angle cut}
\begin{figure}[htbp]
 \begin{center}
    \includegraphics[width=76mm]{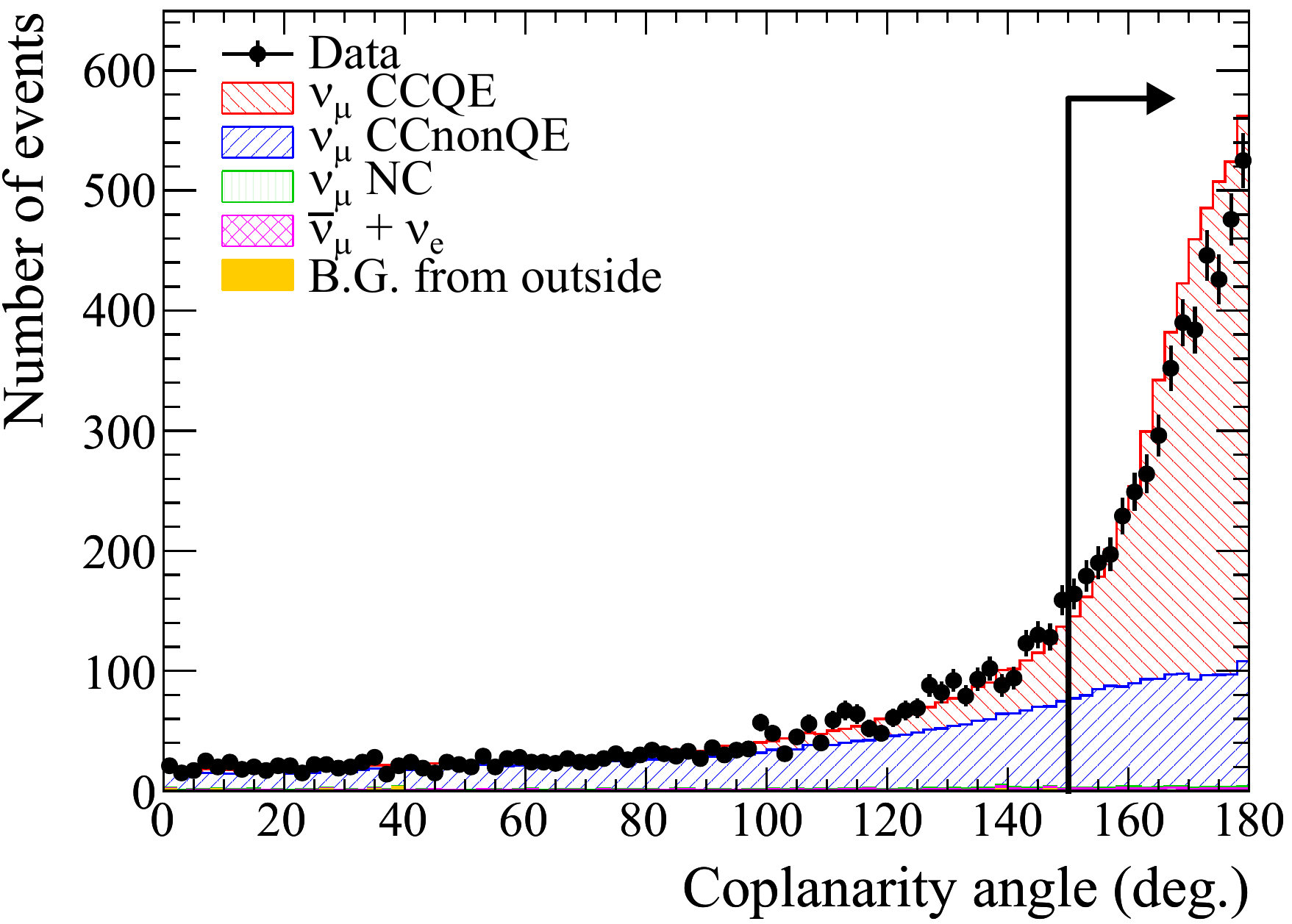}
  \caption{Coplanarity angle distribution following the PID cut for the two track sample.}
\label{coplanarity_ccqe}
   \end{center}
\end{figure}

Since CCQE events are (quasi) two-body scattering interactions, all the tracks in a CCQE event (an incident neutrino track, a scattered muon track and a scattered proton track) are expected to lie in the same plane if the effects of the proton re-scatterings and the Fermi momentum of the target nucleons are neglected.
To quantify this, the coplanarity angle is defined as the angle between the two reconstructed three-dimensional tracks projected to the XY plane, where the XY plane is perpendicular to the neutrino beam axis (Fig.~\ref{kinematic}).
When the three tracks are precisely coplanar, the coplanarity angle is 180$^\circ$. Thus, events with a coplanarity angle above 150$^\circ$ are selected (Fig.~\ref{coplanarity_ccqe}).

\subsubsection{Opening angle cut}
\begin{figure}[htbp]
 \begin{center}
    \includegraphics[width=76mm]{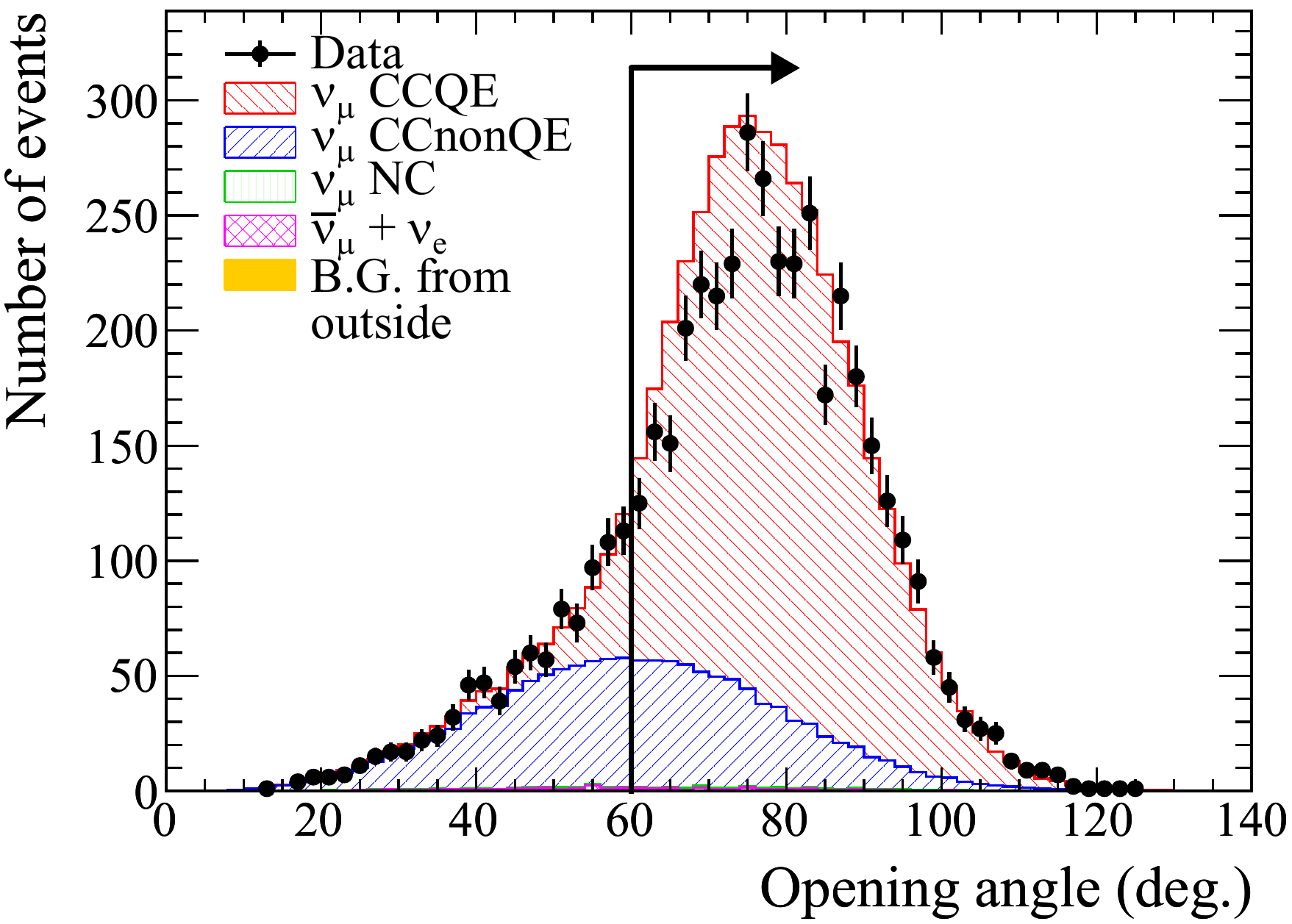}
  \caption{Opening angle distribution following the PID and coplanarity angle cuts for the two track sample.}
\label{opening_ccqe}
   \end{center}
\end{figure}

The opening angle is defined as the angle between the two reconstructed three-dimensional tracks (Fig.~\ref{kinematic}).
The opening angle tends to be large in CCQE interactions, because in the center of mass frame the two final particles are produced back to back. Thus, events with
an opening angle above 60$^\circ$ are selected (Fig.~\ref{opening_ccqe}).
The results of the event selection so far are summarized in Table \ref{event_selection_summary_ccqe}.

\begin{table}[htbp]
	\begin{center}
		\caption{The number of events passing each CCQE selection step. The efficiency is defined as the number of selected $\nu_\mu$ CCQE events divided by the number of $\nu_\mu$ CCQE interactions in the FV. The purity is defined as the ratio of the selected $\nu_\mu$ CCQE events to the total selected events.}
		\vspace*{-0.1cm}
		\begin{tabular}{lrrrr} \hline \hline
			Selection &Data &MC &Efficiency&Purity\\ \hline
			One track from vertex & 12896 & 1.23$\times 10^{4}$& 25.5\%&60.1\%\\
			Particle identification      & 9059 & 8.75$\times 10^{3}$& 23.0\%& 76.1\%\\
			\hline
			Two tracks from vertex &14479  & 1.47$\times 10^{4}$& 20.1\%& 39.4\%\\
			One matched track &13436  & 1.37$\times 10^{4}$& 19.6\%& 41.3\%\\
			Particle identification      &7981  & 8.32$\times 10^{3}$& 15.8\%& 54.9\%\\
			Kinematic cut                &3832  & 4.23$\times 10^{3}$& 12.2\%& 83.5\%\\
			\hline \hline
		\end{tabular}
		\label{event_selection_summary_ccqe}
	\end{center}
\end{table}

\subsection{Energy classification}
\begin{figure}[htbp]
  \begin{center}
  \includegraphics[width=76mm]{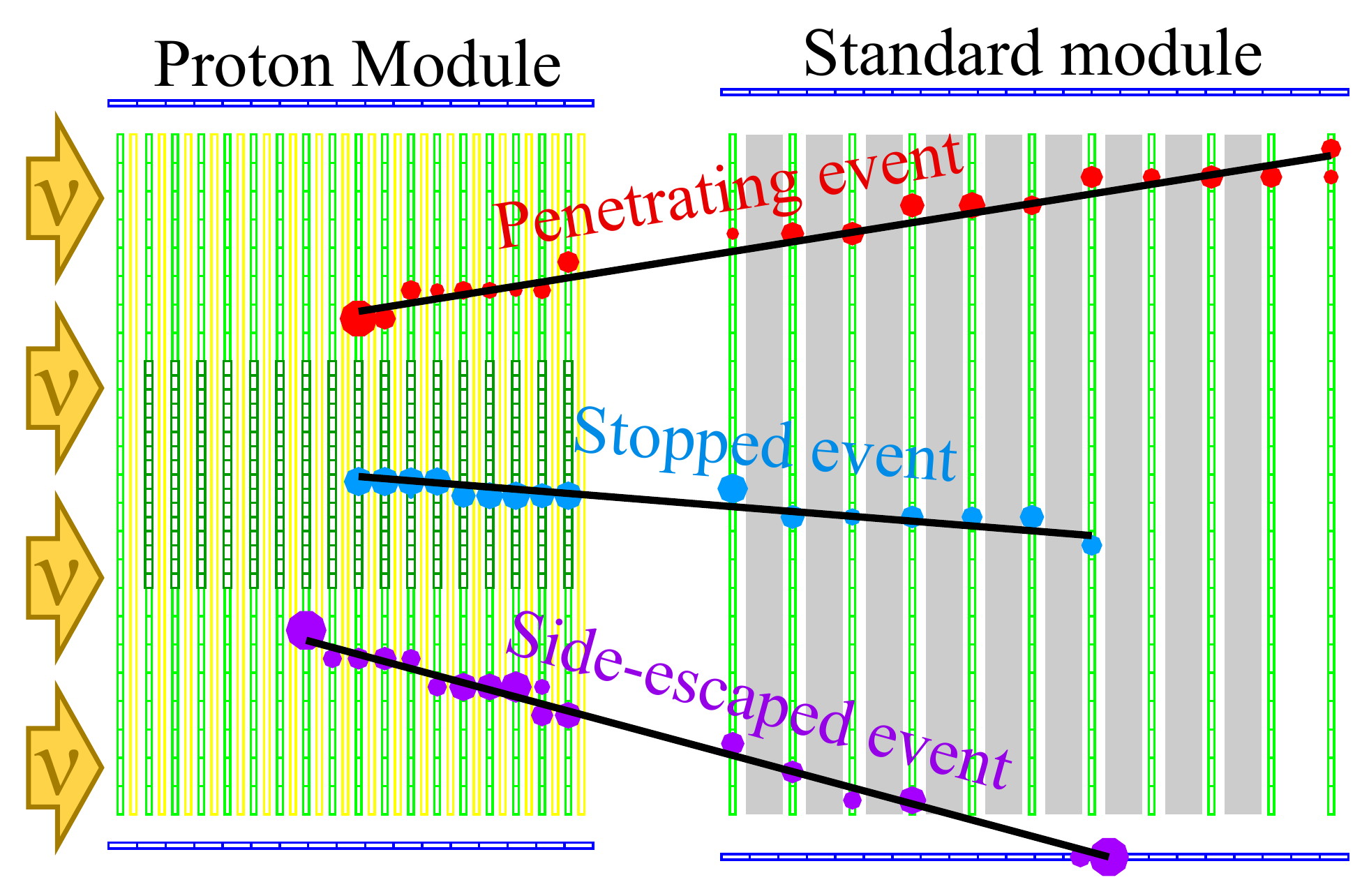}
  \caption{Event display of penetrating, stopped and side-escaped events.}
  \label{energy_selection}
  \end{center}
\end{figure}

\begin{figure}[htbp]
  \begin{center}
  \includegraphics[height=76mm, angle=90]{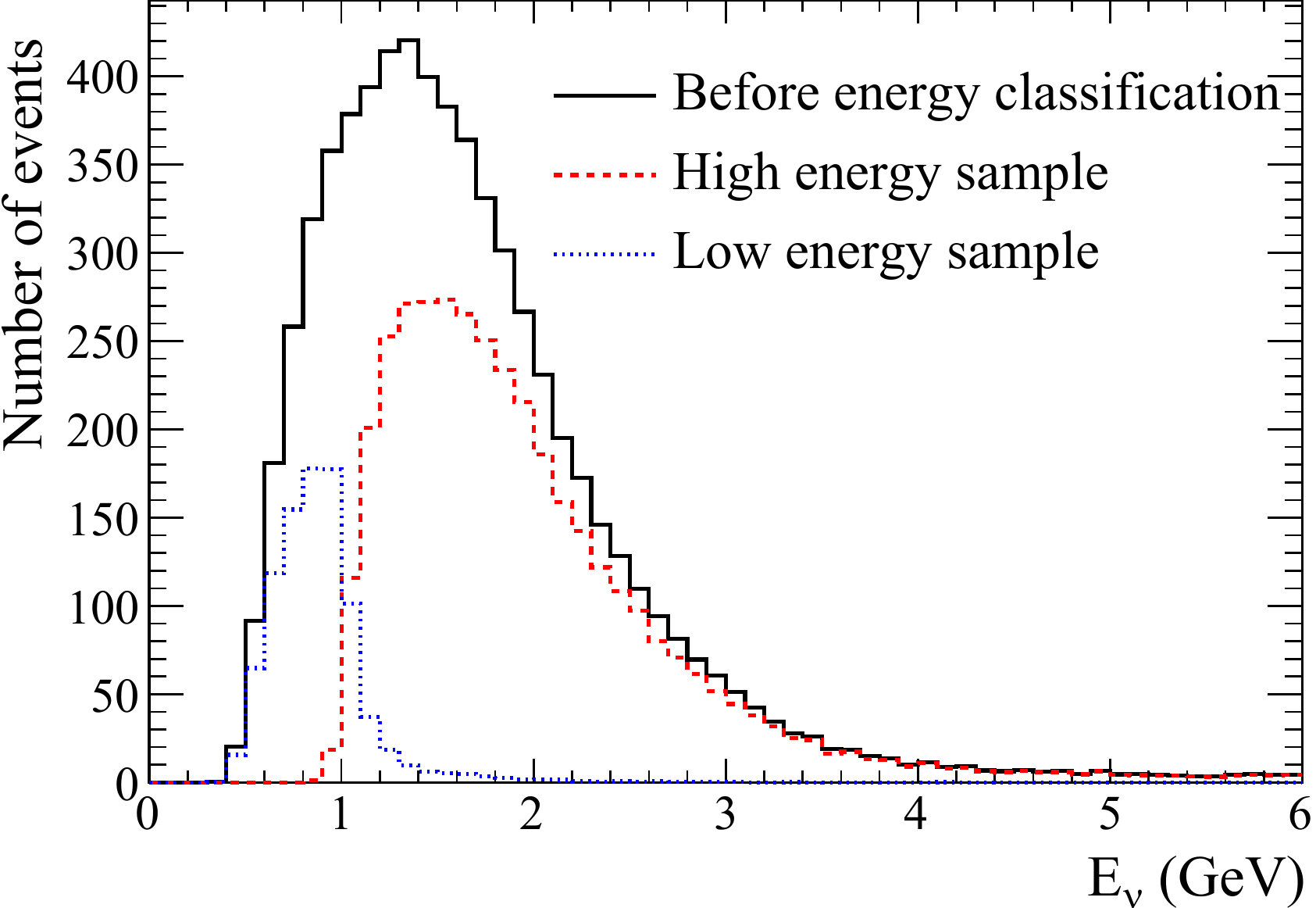}
  \caption{True neutrino energy spectra of the CCQE enhanced sample before and after applying the energy classification for the one-track sample in the MC simulation. The spectrum before the energy classification includes the side-escaped events.}
  \label{enu_1trk}
  \end{center}
\end{figure}

\begin{figure}[htbp]
  \begin{center}
  \includegraphics[height=76mm, angle=90]{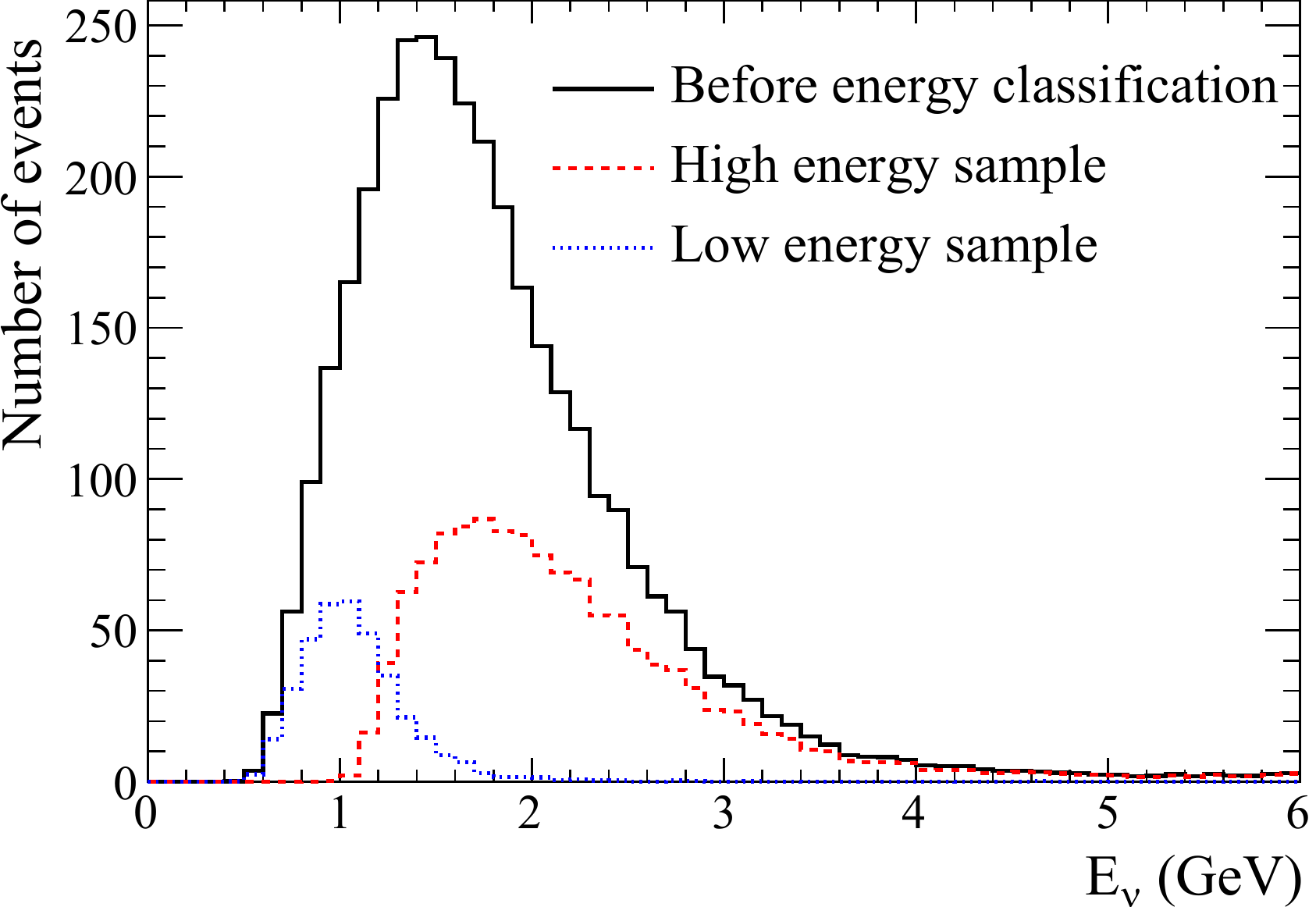}
  \caption{True neutrino energy spectra of the CCQE enhanced sample before and after applying the energy classification for the two-track sample in the MC simulation. The spectrum before the energy classification includes the side-escaped events.}
  \label{enu_2trk}
  \end{center}
\end{figure}

\begin{figure*}[htbp]
	\begin{center}
		\begin{minipage}[b]{.5\linewidth}
			\centering\includegraphics[width=90mm]{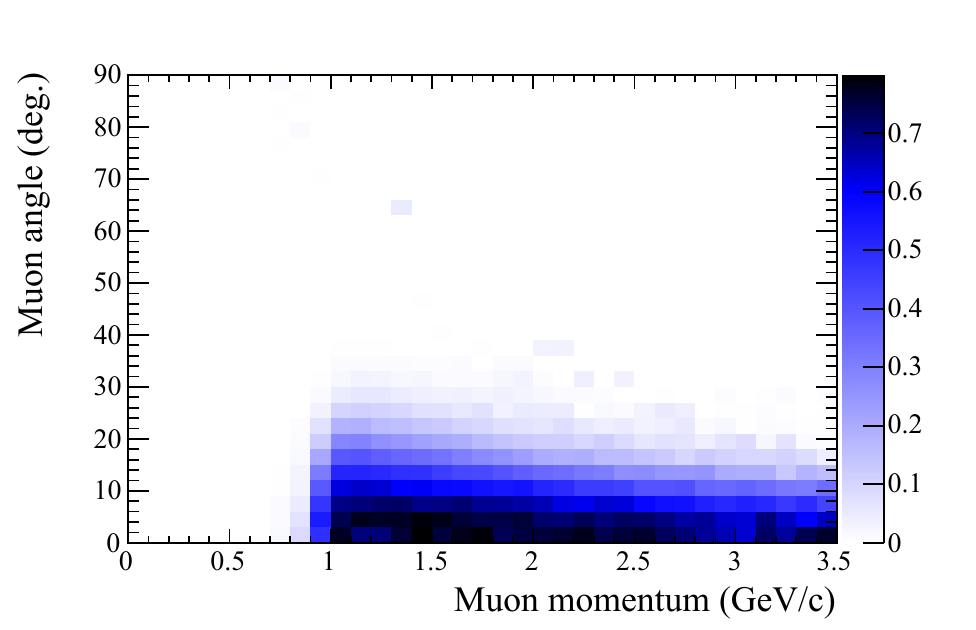}
			\subcaption{High energy one-track sample}\label{eff_ptheta_he1}
		\end{minipage}%
		\begin{minipage}[b]{.5\linewidth}
			\centering\includegraphics[width=90mm]{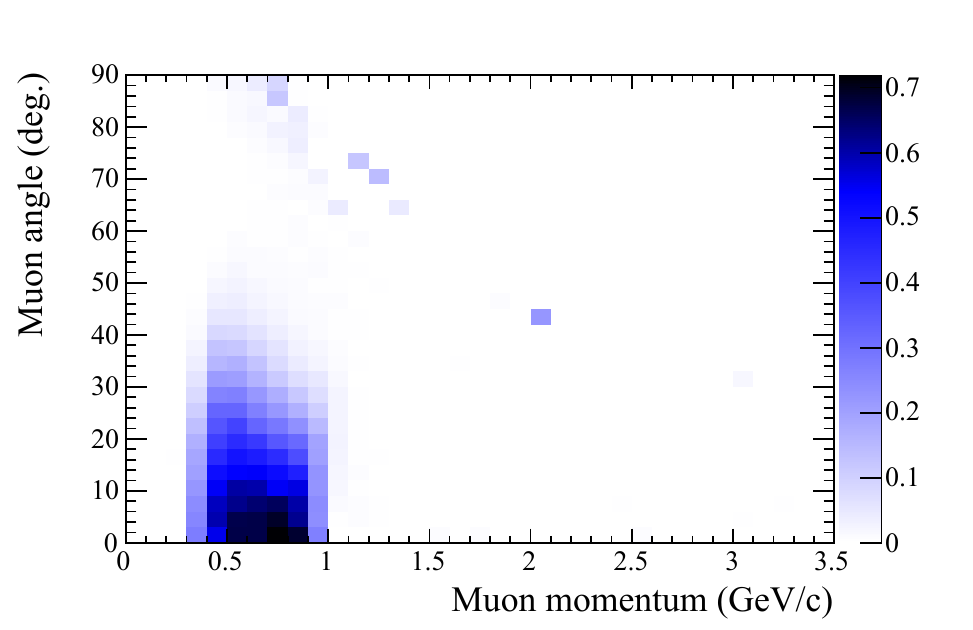}
			\subcaption{Low energy one-track sample}\label{eff_ptheta_le1}
		\end{minipage}
		\begin{minipage}[b]{.5\linewidth}
			\centering\includegraphics[width=90mm]{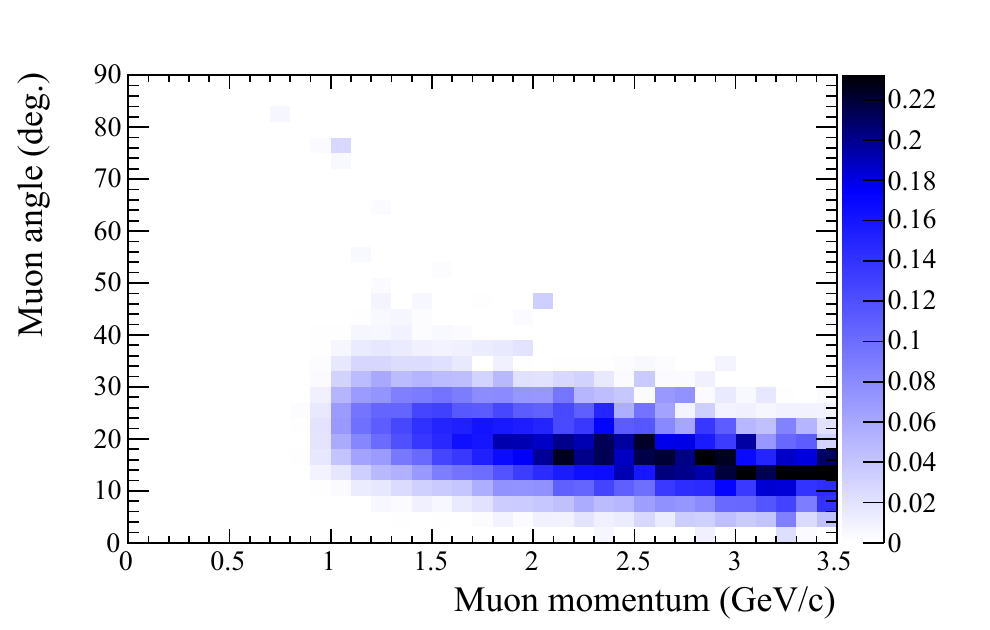}
			\subcaption{High energy two-track sample}\label{eff_ptheta_he2}
		\end{minipage}%
		\begin{minipage}[b]{.5\linewidth}
			\centering\includegraphics[width=90mm]{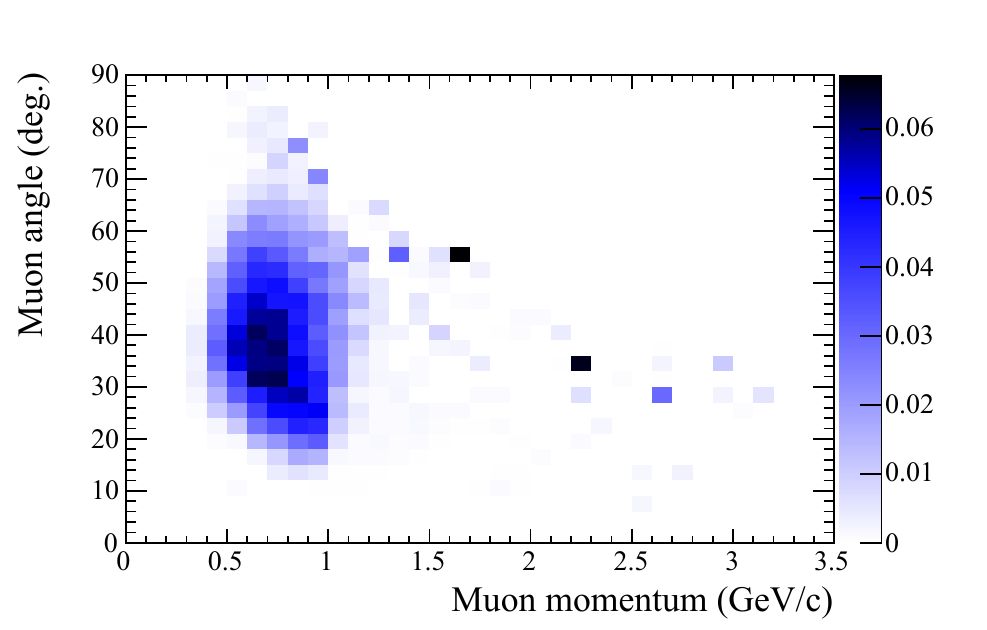}
			\subcaption{Low energy two-track sample}\label{eff_ptheta_le2}
		\end{minipage}
		\caption{Selection efficiency of each sample as a function of the muon momentum and angle.}
		\label{eff_ptheta}
	\end{center}
\end{figure*}

We aim to measure the CCQE cross-section in the low energy region ($\sim$1~GeV) and the high energy region ($\sim$2~GeV) separately.
Thus, an energy classification is applied to the CCQE enhanced samples to select subsamples enriched in high-energy and low-energy events
respectively.
The classification criterion is shown in Fig.~\ref{energy_selection}.
Events with a muon candidate track which penetrates all the standard module iron layers are selected as the high-energy sample, while
events with a muon candidate track which stops in the standard module are selected as the low-energy sample. Other events with a muon candidate track which escapes from the side of the standard module are not used in this analysis.
Figures~\ref{enu_1trk} and \ref{enu_2trk} show the neutrino energy spectra of the CCQE enhanced samples before and after applying the energy classification.
Most of the CCQE events at neutrino energies below 1.0~GeV (above 1.5~GeV) are rejected by the high energy selection (the low energy selection).
Figure~\ref{eff_ptheta} shows the selection efficiency of each sample as a function of the muon momentum and angle. Each sample covers the different muon kinematic region.

\subsection{Event pileup correction}
The T2K neutrino beam is pulsed. Each pulse has an eight-bunch structure, and each bunch has a width of 58~ns.
When a track from a neutrino event piles up with a track from another neutrino event in the same beam bunch, vertices may fail to be reconstructed.
Because this results in the loss of events, this event-pileup effect needs to be corrected for.
The event-pileup effect is proportional to the beam intensity.
Hence, the correction factor is estimated as a linear function of the beam intensity, where
the slope of the linear function is estimated from beam data as follows.
First, the beam data is categorized into subsamples according to the beam intensity.
In each subsample, all hits in INGRID from two beam bunches are summed together to make one new pseudo beam bunch.
This procedure effectively doubles the beam intensity observed by INGRID.
A slope is estimated from the number of selected events in an original beam bunch and a pseudo beam bunch for each subsample.
The slopes estimated from all subsamples are consistent with each other, and the average value of this slope is used for the correction.
This event pileup correction is applied bunch-by-bunch using the slope and POT per bunch which corresponds to the relevant beam intensity.
The event pileup correction gives 0.3--0.7\% difference in the number of selected event in each sample.

\section{Analysis method}\label{sec:method}
\begin{figure}[htbp]
 \begin{center}
  \includegraphics[width=78mm]{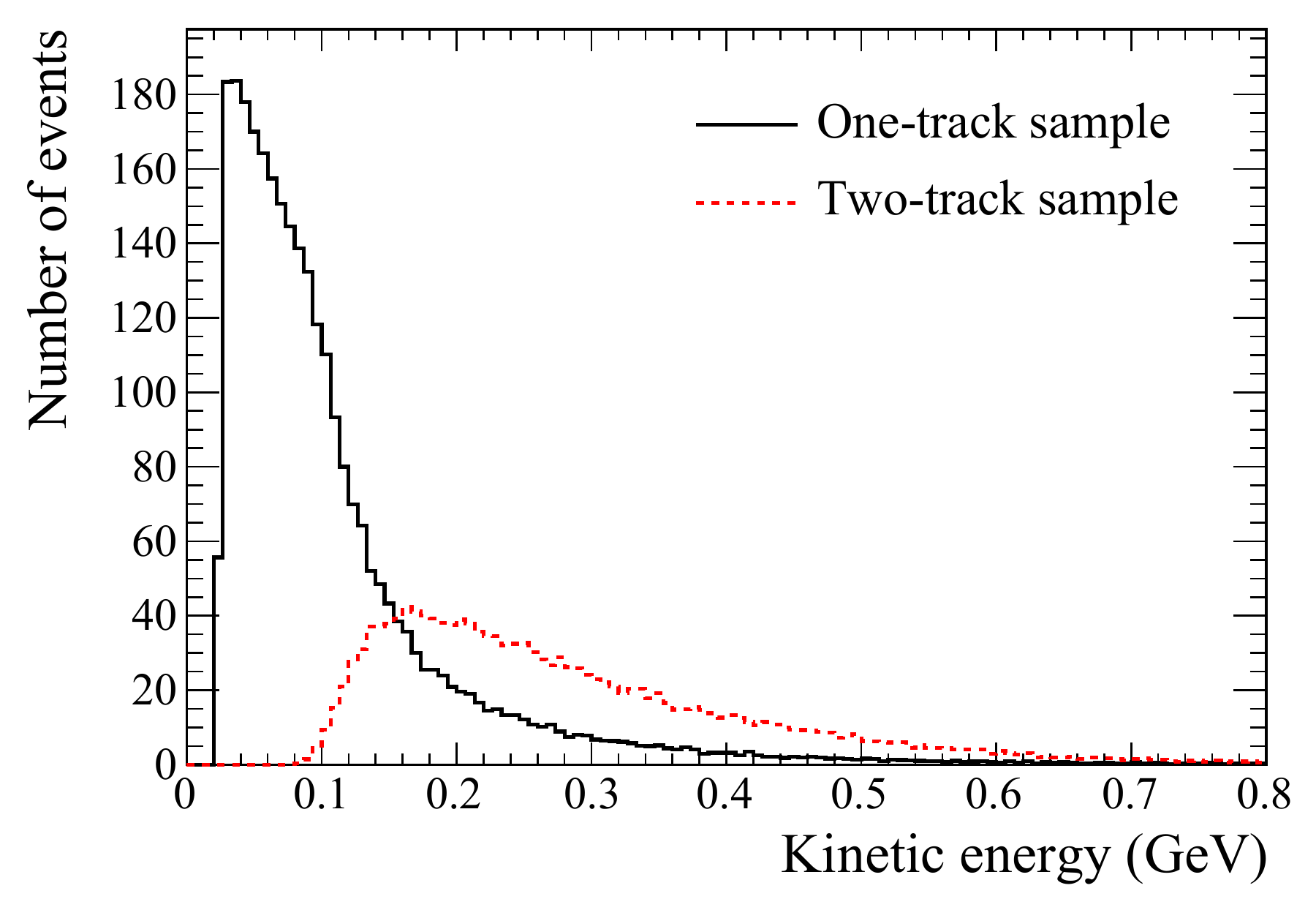}
  \caption{Distributions of the true kinetic energy of protons from the CCQE events in the one-track sample and the two-track sample in the MC simulation.
The cutoff around 0.02~GeV stems from the effect of Pauli blocking.}
  \label{t_1_2_track}
   \end{center}
\end{figure}

We estimate the CCQE cross-sections in high and low energy regions, which are defined as above 1.0~GeV and below 1.5~GeV using the high and low energy samples.
The average energies of the neutrino flux in the high and low energy regions are 1.94~GeV and 0.93~GeV, respectively. 
The CCQE cross-section is calculated from the number of selected CCQE candidate events by subtracting background and correcting for selection efficiency:
\begin{equation}
\sigma_{\mathrm{CCQE}} = \frac{N_{\mathrm{sel}}-N_{\mathrm{BG}}}{\phi T\varepsilon},\label{eq_ccqe}
\end{equation}
where $N_{\mathrm{sel}}$ is the number of selected CCQE candidate events from real data, $N_{\mathrm{BG}}$ is the number of selected background events predicted by the MC simulation, $\phi$ is the integrated $\nu_{\mu}$ flux,  $T$ is the number of target neutrons, and  $\varepsilon$ is the detection efficiency for CCQE events predicted by the MC simulation.
The flux $\phi$ is integrated in each energy region, and $\varepsilon$ is calculated for CCQE events in each energy region.
CCQE events assigned to the wrong subsample, i.e.\ high-energy events in the low-energy subsample and vice versa, are regarded as background.
The background events for this analysis consist of CC-nonQE events, NC events, $\bar{\nu}_{\mu}$ events, $\nu_{e}$ events, external background events created by neutrino interactions in the material surrounding the detector, and CCQE events assigned to the wrong energy sample.
Furthermore, the CCQE cross-section in each energy region is estimated from the one-track sample, two-track sample, and combined sample, separately.
The one-track sample (the two-track sample) has an enhanced content of low (high) energy protons from CCQE interactions as shown in Fig.\ \ref{t_1_2_track}.
Therefore, we can cross-check the CCQE cross-section results from the different phase spaces.

\section{Systematic errors}\label{sec:error}
Uncertainties on $N_{\mathrm{BG}}$, $\phi$, $T$, and $\varepsilon$ are sources of systematic errors on the cross-section results.
The sources of systematic error can be categorized into three groups: those from the neutrino flux prediction, the neutrino interaction model including intra-nuclear interactions, and the detector response.

\subsection{Neutrino flux uncertainties}
The neutrino flux uncertainty sources can be separated into two categories: hadron production uncertainties and T2K beamline uncertainties.
The uncertainties on hadron production are mainly
driven by the NA61/SHINE measurements \cite{na61_1, na61_2} and the
Eichten and Allaby data \cite{eichten, allaby}, and constitute the dominant
component of the flux uncertainty.
They include the uncertainties on the production cross-section, the secondary nucleon production, the pion production multiplicity, and the kaon
production multiplicity.
The second category of flux uncertainties is associated
with inherent uncertainties and operational variations in the beamline conditions, including
 uncertainties in the proton beam position, the beam-axis direction, the absolute horn
current, the horn angular alignment, the horn field
asymmetry, the target alignment, and the proton beam intensity.
The method of estimating these flux uncertainties is described in Ref.\  \cite{flux_prediction}.
To evaluate the systematic error from the flux uncertainties, the flux is varied using a covariance matrix based on the flux uncertainty.
This is repeated for many toy data sets, and the $\pm$1$\sigma$ of the change in the cross-section result is taken as the systematic error associated with the neutrino flux.
The systematic error is 11-17\%, which is the dominant error in this measurement.

\subsection{Neutrino interaction uncertainties}
We use a data-driven method to calculate the neutrino interaction uncertainties, where the NEUT predictions are compared to available external neutrino-nucleus data in the energy region relevant for T2K.
We fit some parameters of the models implemented in NEUT, and introduce ad hoc parameters, often with large uncertainties, to take into account remaining discrepancies between NEUT and the external data \cite{mb_ccpi0, mb_ccpipm, mb_ncpi0, miniboone_ccqe, nomad_ccqe, minerva_ccqe_nu, minerva_ccqe_nubar, minerva_cc1pi, k2k_coh, sciboone_coh, sciboone_nccoh, minos_ccinc}.

The model parameters include axial mass values for quasi-elastic scattering and meson production via baryon resonances, the Fermi momentum, the binding energy, a spectral function parameter, and a $\pi$-less $\Delta$ decay parameter.
The spectral function parameter is introduced to take into account the difference between the relativistic Fermi gas nuclear model, which is the standard NEUT model, and the more sophisticated spectral function model \cite{sf_imp}, which is expected from electron scattering data \cite{escat1, escat2} to be a better representation of the nuclear model.
The implemented ad hoc parameters include cross-section normalizations.
In addition, uncertainties on the final state interactions of the pions and nucleons with the nuclear medium are included.
Table~\ref{int_par_uncertainty} shows the nominal values and uncertainties on these parameters.
Systematic errors from the nuclear model (spectral function, Fermi momentum and binging energy) were found to be comparatively large because the CCQE interaction in the few GeV region is sensitive to the nuclear model.
Further details about these uncertainties are described in Refs \cite{t2k_ccinc,t2k_nue_2013}.
Systematic errors due to these parameters are estimated from variations of the cross-section results when these parameters are varied within their uncertainties.
As a cross check, we also estimated the selection efficiencies using the GENIE neutrino interaction generator and confirmed that they were consistent with those using NEUT when the axial vector mass is set to the same value as GENIE, 0.99~GeV.

\begin{table}[htbp]
\begin{center}
\caption{The nominal values of and the uncertainties on the neutrino interaction model parameters. The first, second, and third groups represent the model parameters, the ad hoc parameters, and the final state interaction parameters respectively. The $\pi$-less $\Delta$ decay parameter and the final state interaction parameters vary the probabilities of these interactions. Nominal values of 0 or 1 mean that the effect or the normalization is not implemented or is implemented by default, respectively.}
\begin{tabular}{ccc} \hline \hline
Parameter &  Nominal & Error\\ \hline
$M_A^{\mathrm{QE}}$	&	1.21GeV	&	16.53\%	\\
$M_A^{\mathrm{RES}}$	&	1.21GeV	&	16.53\%	\\
$\pi$-less $\Delta$ decay &0.2 & 20\%\\
Spectral function	&	0(off)	&	100\%	\\
Fermi momentum &	217MeV/c	&	13.83\%	\\
Binding energy &	25MeV	&	36\%	\\
\hline
CC1$\pi$ norm.  ($E_{\nu}<$2.5GeV)	&	1	&	21\%	\\
CC1$\pi$ norm.  ($E_{\nu}>$2.5GeV)	&	1	&	21\%	\\
CC coherent $\pi$ norm.	&	1	&	100\%	\\
CC other shape	&	0(off)	&	40\% at 1~GeV\\
NC1$\pi^0$ norm.	&	1	&	31\%	\\
NC coherent $\pi$ norm.	&	1	&	30\%	\\
NC1$\pi^{\pm}$ norm.	&	1	&	30\%	\\
NC other norm.	&	1	&	30\%	\\
\hline
$\pi$ absorption	&	1	&	50\%	\\
$\pi$ charge exchange (low energy)	&	1	&	50\%	\\
$\pi$ charge exchange (high energy)	&	1	&	30\%	\\
$\pi$ QE scattering (low energy)	&	1	&	50\%	\\
$\pi$ QE scattering (high energy)	&	1	&	30\%	\\
$\pi$ inelastic scattering	&	1	&	50\%	\\
Nucleon elastic scattering & 1 & 10\% \\
Nucleon single $\pi$ production & 1 & 10\% \\
Nucleon two $\pi$ production & 1 & 10\% \\
\hline \hline
\end{tabular}
\label{int_par_uncertainty}
  \end{center}
\end{table}

\subsection{Detector response uncertainties}
The uncertainty of the target mass measurement, 0.25\%, is taken as the systematic error on the target mass.
The variation in the measured MPPC dark rate during data acquisition, 11.52 hits/cycle, where `cycle' denotes the integration cycle synchronized with the neutrino beam pulse structure, is taken as the uncertainty on the MPPC dark rate.
The discrepancy between the hit detection efficiency measured with beam-induced muon backgrounds and that of the MC simulation, 0.21\%, is assigned as the uncertainty in the hit detection efficiency.
The uncertainty of the light yield is evaluated by using the beam-induced muon backgrounds as a control sample; in addition, the uncertainty of the scintillator quenching is taken into account based on the uncertainty of the Birks' constant (0.0208$\pm$0.0023 cm/MeV).
The relations between these quantities and the cross-section results are estimated by MC simulation, and the resulting variations on the calculated cross-sections are assigned as systematic errors.
The event pileup correction factor has uncertainties which come from the statistics of the beam data and the MPPC dark count in the estimation of the correction factor.
The systematic error from these uncertainties is estimated assuming the highest beam intensity achieved in beam operation so far.
There is about a 35\% discrepancy between the beam-induced muon background rate estimated by the MC simulation and that measured from the data.
The change in the background contamination fraction from this discrepancy is taken as the systematic error for the beam-related background.
The cosmic-ray background was found to be very small from the out-of-beam timing data.
The systematic error on the track reconstruction efficiency is estimated by comparing the efficiency for several subsamples between the data and the MC simulation.
The standard deviation of the difference of the track reconstruction efficiency between data and MC for the subsamples is taken as the systematic error.
The systematic errors from all event selections are evaluated by varying each selection threshold.
The maximum difference between the data and MC for each selection threshold is taken as the value of each systematic error.
Among the systematic errors from the detector response, the largest contributions (about 2\% each) are those from the light yield and the secondary interactions.

\subsection{Summary of the systematic errors}
The total systematic errors, calculated from the quadrature sum of all the systematic errors,
on the CCQE cross-section from the one-track sample, the two-track sample and the combined sample for the high energy region are $_{-12.97\%}^{+15.95\%}$, $_{-14.04\%}^{+16.97\%}$ and $_{-12.44\%}^{+15.06\%}$, respectively, and those for the low energy region are $_{-17.04\%}^{+20.35\%}$, $_{-18.86\%}^{+24.20\%}$ and $_{-15.49\%}^{+19.04\%}$, respectively.
Table~\ref{ccqe_syst} summarizes the breakdown of the systematic errors on the CCQE cross-section measurement from the combined sample.
The neutrino flux error is the dominant systematic error.

\begin{table*}[htbp]
\begin{center}
\caption{Summary of the systematic errors on the CCQE cross-section measurement from the combined sample. Negative and positive values represent $-1\sigma$ and $+1\sigma$ errors.}
\begin{tabular}{lcc}
\hline \hline
Item &High energy region&Low energy region\\ \hline
Neutrino flux&$-11.01\%+13.61\%$&$-13.57\%+17.04\%$\\
\hline
$M_A^{\mathrm{QE}}$&$-0.89\%+2.25\%$&$-0.08\%+0.39\%$\\
$M_A^{\mathrm{RES}}$&$-0.92\%+1.31\%$&$-0.82\%+1.10\%$\\
CC1$\pi$ normalization  ($E_{\nu}<$2.5~GeV)&$-0.55\%+0.50\%$&$-3.71\%+3.59\%$\\
CC1$\pi$ normalization  ($E_{\nu}>$2.5~GeV)&$-2.69\%+2.69\%$&$-1.88\%+1.83\%$\\
CC coherent $\pi$ normalization&$-1.40\%+1.38\%$&$-1.73\%+1.71\%$\\
CC other $E_{\nu}$ shape&$-0.86\%+0.85\%$&$-0.11\%+0.09\%$\\
NC1$\pi^0$ normalization&$-0.65\%+0.65\%$&$-0.40\%+0.40\%$\\
NC coherent $\pi$ normalization&$-0.10\%+0.10\%$&$-0.09\%+0.09\%$\\
NC1$\pi^{\pm}$ normalization&$-0.47\%+0.47\%$&$-0.46\%+0.45\%$\\
NC other normalization&$-0.33\%+0.31\%$&$-0.75\%+0.74\%$\\
$\pi$-less $\Delta$ decay&$-0.54\%+2.10\%$&$-1.60\%+3.34\%$\\
Spectral function&$-2.01\%+0.00\%$&$-0.00\%+1.21\%$\\
Fermi momentum&$-1.67\%+2.22\%$&$-3.71\%+4.43\%$\\
Binding energy&$-0.44\%+0.65\%$&$-1.24\%+1.42\%$\\
Pion absorption&$-0.20\%+0.81\%$&$-0.80\%+1.20\%$\\
Pion charge exchange (low energy)&$-0.15\%+0.18\%$&$-0.22\%+0.28\%$\\
Pion charge exchange (high energy)&$-0.11\%+0.13\%$&$-0.11\%+0.11\%$\\
Pion QE scattering (low energy)&$-0.66\%+0.71\%$&$-0.84\%+0.79\%$\\
Pion QE scattering (high energy)&$-0.04\%+0.03\%$&$-0.09\%+0.09\%$\\
Pion inelastic scattering&$-0.05\%+0.04\%$&$-0.29\%+0.25\%$\\
Nucleon elastic scattering&$-0.25\%+0.21\%$&$-0.29\%+0.21\%$\\
Nucleon single $\pi$ production &$-0.15\%+0.11\%$&$-0.60\%+0.51\%$\\
Nucleon two $\pi$ production &$-0.57\%+0.42\%$&$-0.01\%+0.01\%$\\
\hline
Target mass&$\pm$0.31\%&$\pm$0.38\%\\
MPPC dark noise&$\pm$0.03\%&$\pm$0.08\%\\
Hit efficiency&$\pm$0.84\%&$\pm$0.41\%\\
Light yield&$\pm$1.47\%&$\pm$2.22\%\\
Event pileup&$\pm$0.02\%&$\pm$0.06\%\\
Beam-induced external background&$\pm$0.08\%&$\pm$0.35\%\\
Cosmic-ray background&$\pm$0.00\%&$\pm$0.01\%\\
2D track reconstruction&$\pm$0.67\%&$\pm$0.81\%\\
Track matching&$\pm$0.45\%&$\pm$1.13\%\\
3D tracking&$\pm$0.21\%&$\pm$0.15\%\\
Vertexing&$\pm$0.30\%&$\pm$0.43\%\\
Timing cut&$\pm$0.00\%&$\pm$0.00\%\\
Veto cut&$\pm$0.82\%&$\pm$0.64\%\\
Fiducial volume cut&$\pm$1.55\%&$\pm$0.84\%\\
Secondary interaction&$\pm$2.45\%&$\pm$2.37\%\\
\hline
Total&$-12.44\%+15.06\%$&$-15.49\%+19.04\%$\\
\hline \hline
\end{tabular}
\label{ccqe_syst}
  \end{center}
\end{table*}

\section{Results}\label{sec:result}
\subsection{Results with the T2K default interaction model}
\begin{figure}[htbp]
  \begin{center}
  \includegraphics[width=80mm]{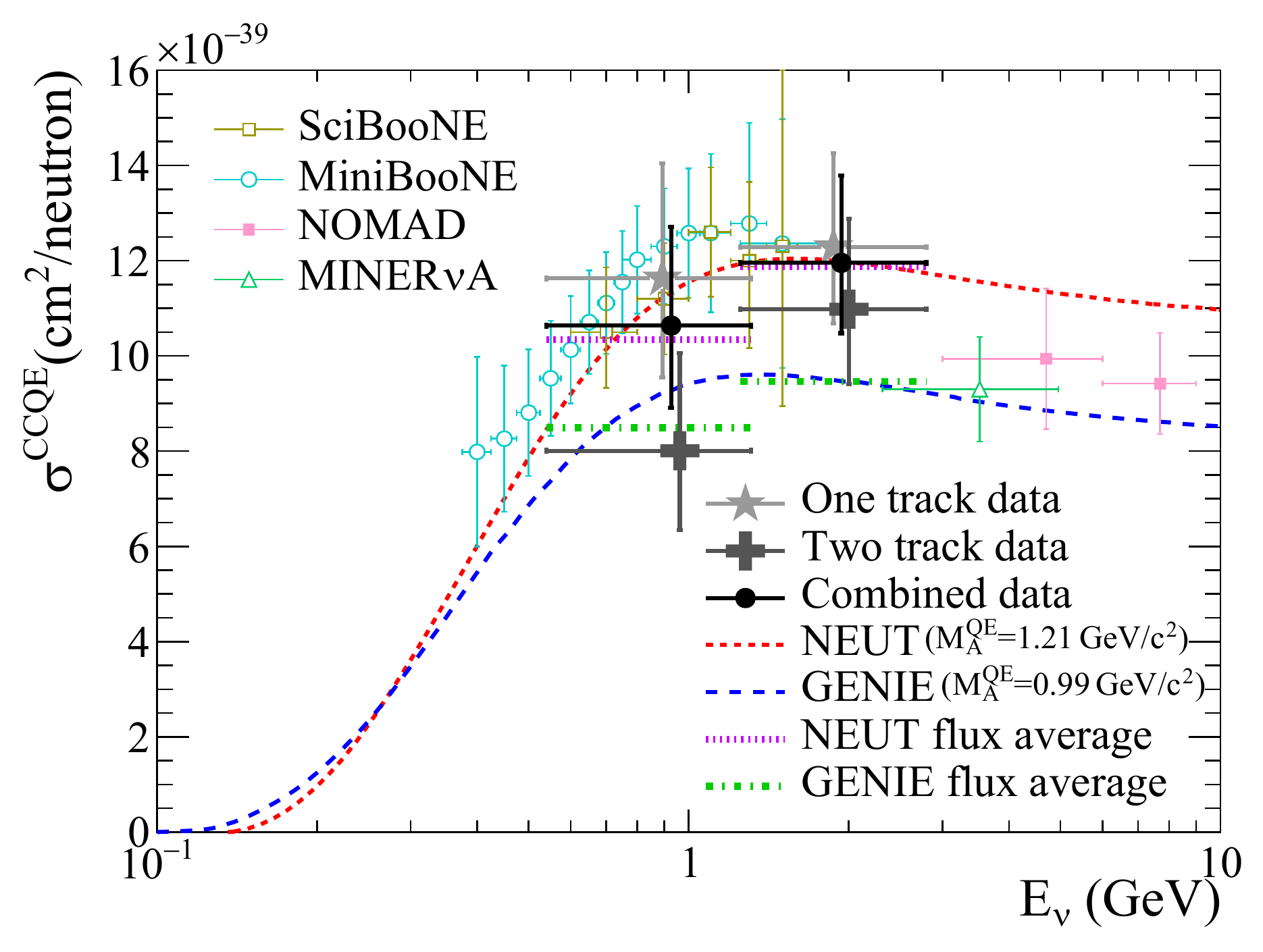}
  \caption{The CCQE cross-section results with predictions by NEUT and GENIE. Our data point is placed at the flux mean energy. The vertical error bar represents the total (statistical and systematic) uncertainty,
and the horizontal bar represent 68\% of the flux at each side of the mean energy. The SciBooNE, MiniBooNE, NOMAD and MINER$\nu$A results are also plotted \cite{sciboone_ccqe, miniboone_ccqe, nomad_ccqe, minerva_ccqe_nu}.}
  \label{ccqe_result}
  \end{center}
\end{figure}

The results of the CCQE cross-section measurement from the different subsamples are summarized in Table \ref{table_ccqe_result} (a).
The measured CCQE cross-sections from the combined sample are
\begin{eqnarray}
\sigma_{\mathrm{CCQE}}(1.94\mathrm{GeV}) &=& (11.95\pm 0.19(stat.)_{-1.49}^{+1.80}
(syst.))\nonumber\\
&\times& 10^{-39}\mathrm{cm}^2/\mathrm{neutron}\\
\sigma_{\mathrm{CCQE}}(0.93\mathrm{GeV}) &=& (10.64\pm 0.37(stat.)_{-1.65}^{+2.03}
(syst.))\nonumber\\
&\times& 10^{-39}\mathrm{cm}^2/\mathrm{neutron},
\end{eqnarray}
at mean neutrino energies of 1.94~GeV and 0.93~GeV (for the high and low energy regions), respectively.
We quote these values as our primary result.
The NEUT and GENIE predictions of the CCQE cross-sections on carbon for the high and low energy regions are shown in Table~\ref{ccqe_prediction}. The difference in the predictions from NEUT and GENIE is attributable primarily to the difference in the nominal $M_A^{\mathrm{QE}}$ value.
The results of the measurements are consistent within 2$\sigma$ with both predictions;
however, in the low-energy region the cross-section results from the one-track and two-track samples differ by just under 2$\sigma$, as shown in Table \ref{ccqe_1trk_2trk_ratio}. 
The cross-section results are shown in Fig.\ \ref{ccqe_result} together with the predictions and the measurements of other experiments, and the reconstructed muon angle distribution of each CCQE candidate sample is shown in Fig.\ \ref{ccqe_muon_angle}. The distributions of other kinematic variables are summarized in the supplemental material \cite{suppl_material}.

\begin{table*}[htbp]
\begin{center}
\caption{The CCQE cross-sections measured from each sample ($\times 10^{-39}$cm$^2$). 
The mean neutrino energies of the high and low energy regions are 1.94\,GeV and 0.93\,GeV, respectively.}
  \vspace*{-0.1cm}
\begin{tabular}{lrr}
\multicolumn{3}{c}{(a) Results with the relativistic Fermi gas model (T2K default interaction model).}\\
\hline \hline

Used sample & High energy region & Low energy region\\ \hline
One-track sample &  $12.29\pm0.22(stat.)_{-1.59}^{+1.96}(syst.)$&  $11.63\pm0.45(stat.)_{-1.98}^{+2.37}(syst.)$\\
Two-track sample &  $10.98\pm0.35(stat.)_{-1.54}^{+1.86}(syst.)$&  $8.01\pm0.64(stat.)_{-1.51}^{+1.94}(syst.)$\\
Combined sample &  $11.95\pm0.19(stat.)_{-1.49}^{+1.80}(syst.)$&  $10.64\pm0.37(stat.)_{-1.65}^{+2.03}(syst.)$\\
\hline \hline
& \\
\multicolumn{3}{c}{(b) Results with the spectral function model.}\\
\hline \hline
Used sample & High energy region & Low energy region\\ \hline
One-track sample & $12.46\pm0.22(stat.)_{-1.62}^{+1.98}(syst.)$ & $11.04\pm0.43(stat.)_{-1.84}^{+2.26}(syst.)$\\
Two-track sample & $11.43\pm0.36(stat.)_{-1.60}^{+1.85}(syst.)$ & $8.84\pm0.70(stat.)_{-1.70}^{+1.94}(syst.)$\\
Combined sample & $12.19\pm0.19(stat.)_{-1.50}^{+1.84}(syst.)$ & $10.51\pm0.37(stat.)_{-1.67}^{+2.00}(syst.)$\\
\hline \hline
& \\
\multicolumn{3}{c}{(c) Results on assuming the multi-nucleon interactions.}\\
\hline \hline
Used sample & High energy region & Low energy region\\ \hline
One-track sample & $10.79\pm0.22(stat.)_{-1.63}^{+2.01}(syst.)$& $10.11\pm0.45(stat.)_{-2.03}^{+2.41}(syst.)$\\
Two-track sample & $10.28\pm0.35(stat.)_{-1.52}^{+1.85}(syst.)$& $7.14\pm0.64(stat.)_{-1.56}^{+1.96}(syst.)$\\
Combined sample & $10.66\pm0.19(stat.)_{-1.52}^{+1.88}(syst.)$& $9.30\pm0.37(stat.)_{-1.75}^{+2.13}(syst.)$\\
\hline \hline

\end{tabular}
\label{table_ccqe_result}
\end{center}
\end{table*}

\begin{table}[htbp]
\begin{center}
\caption{The NEUT and GENIE predictions of the flux-averaged CCQE cross-sections on carbon for the high and low energy regions.}
  \vspace*{-0.1cm}
\begin{tabular}{lrr} \hline \hline
 & High energy region & Low energy region\\ \hline
NEUT & 11.88$\times 10^{-39}$cm$^2$ & 10.34$\times 10^{-39}$cm$^2$\\
GENIE & 9.46$\times 10^{-39}$cm$^2$ & 8.49$\times 10^{-39}$cm$^2$\\ \hline \hline
\end{tabular}
\label{ccqe_prediction}
  \end{center}
\end{table}

\begin{figure*}[htbp]
\begin{center}
\begin{minipage}[b]{.5\linewidth}
	\centering\includegraphics[angle=90, height=90mm]{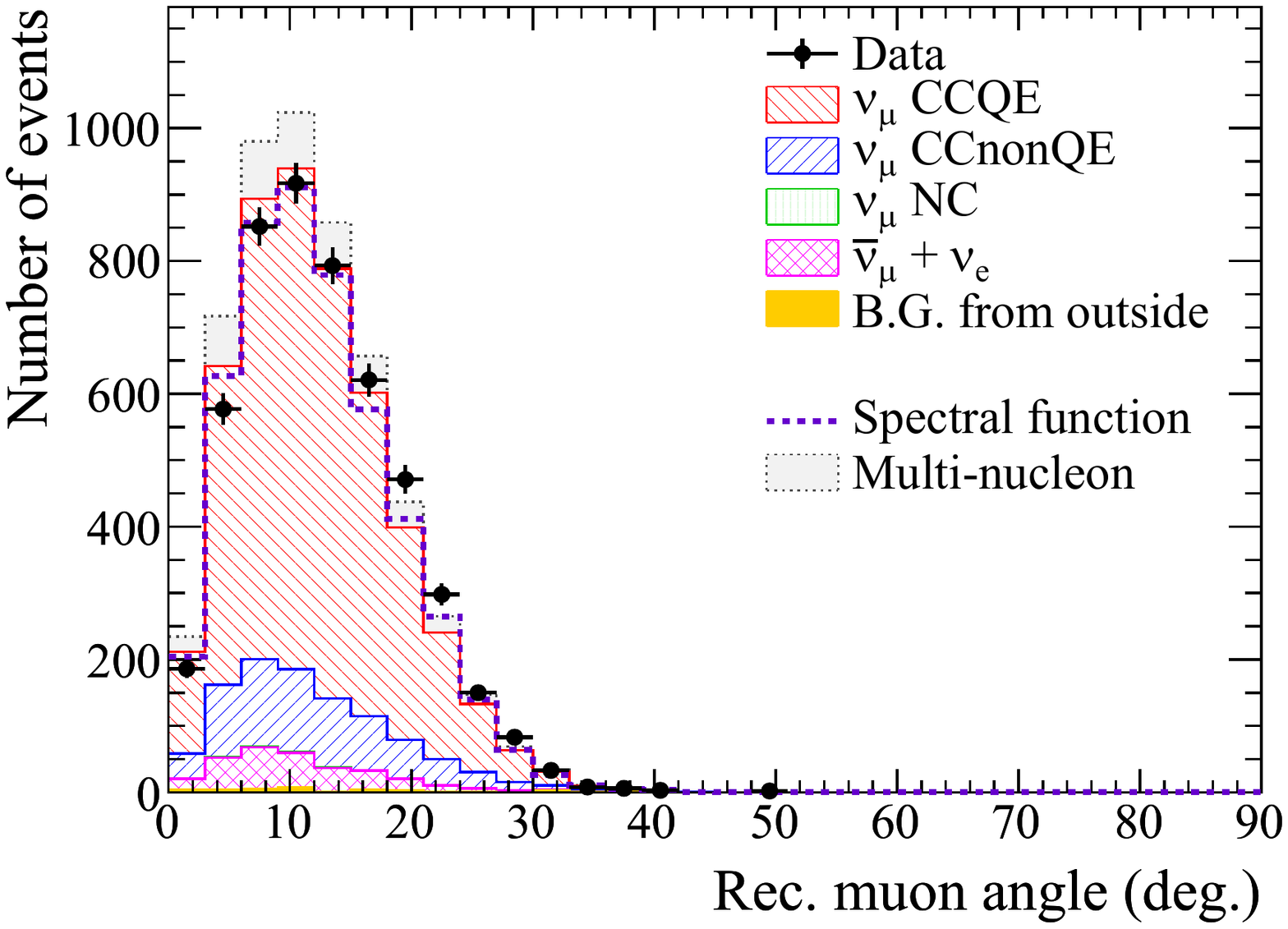}
	\subcaption{High energy one-track sample}\label{muang_he1}
\end{minipage}%
\begin{minipage}[b]{.5\linewidth}
	\centering\includegraphics[angle=90, height=90mm]{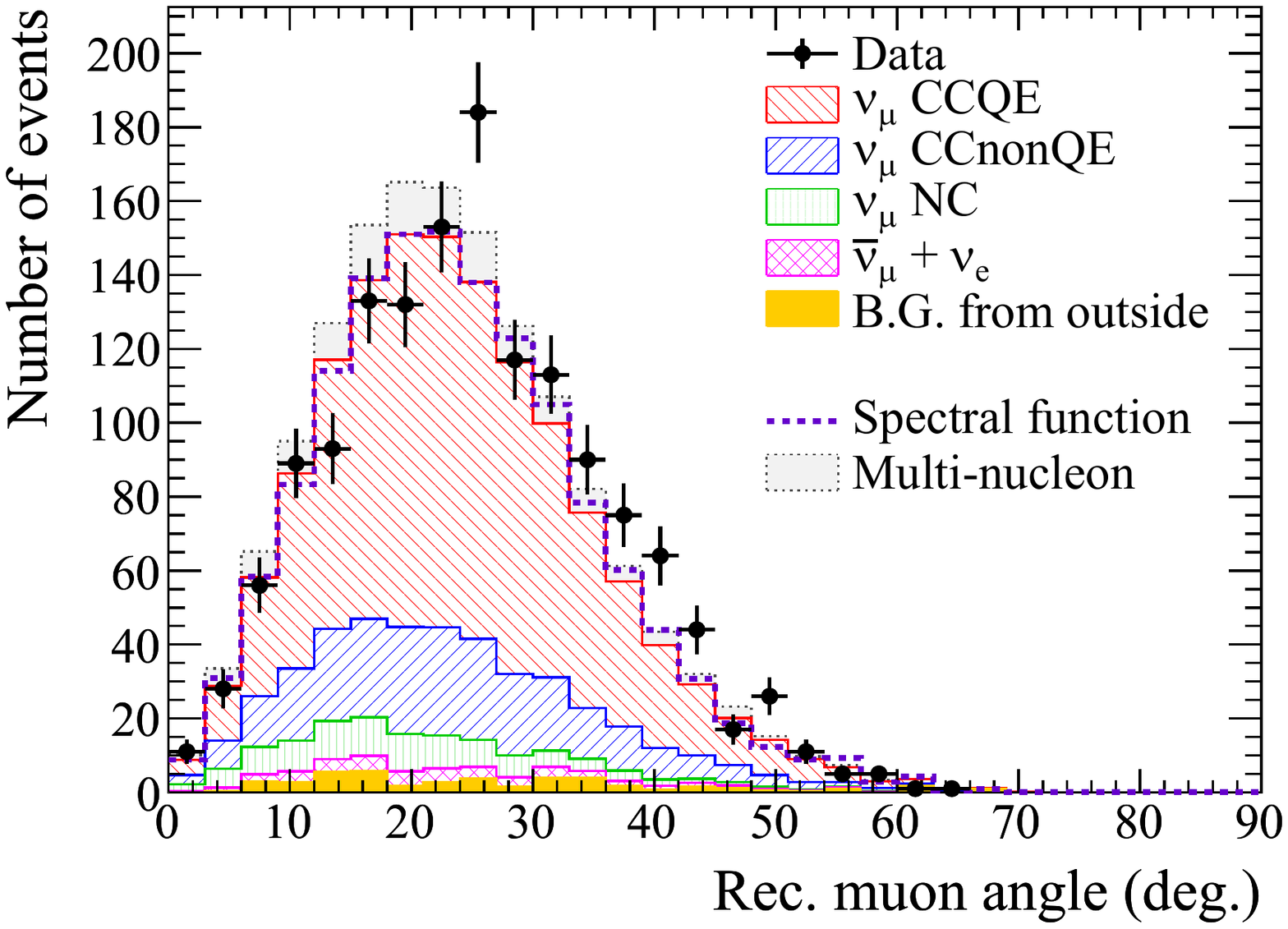}
	\subcaption{Low energy one-track sample}\label{muang_le1}
\end{minipage}
\begin{minipage}[b]{.5\linewidth}
	\centering\includegraphics[angle=90, height=90mm]{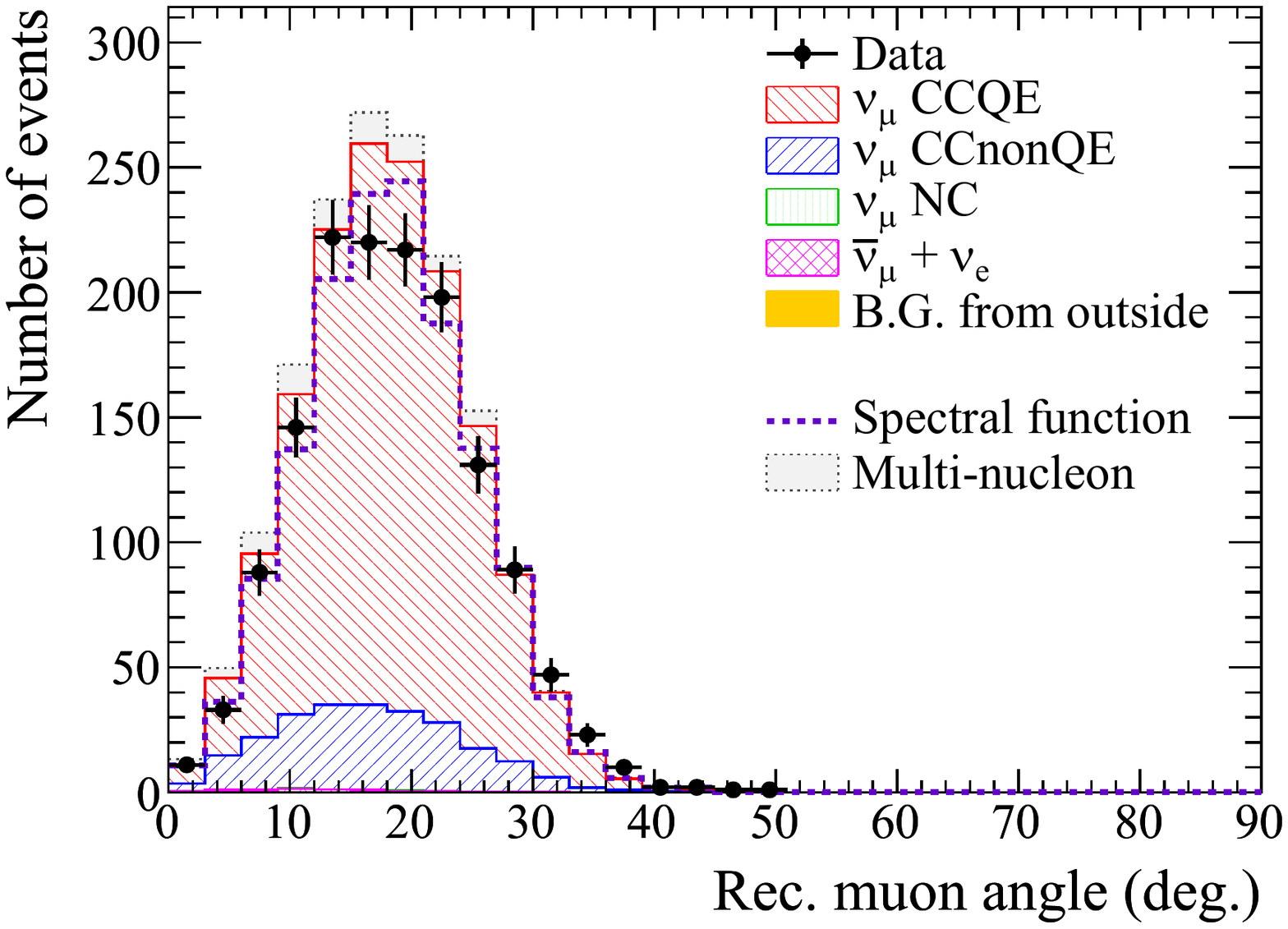}
	\subcaption{High energy two-track sample}\label{muang_he2}
\end{minipage}%
\begin{minipage}[b]{.5\linewidth}
	\centering\includegraphics[angle=90, height=90mm]{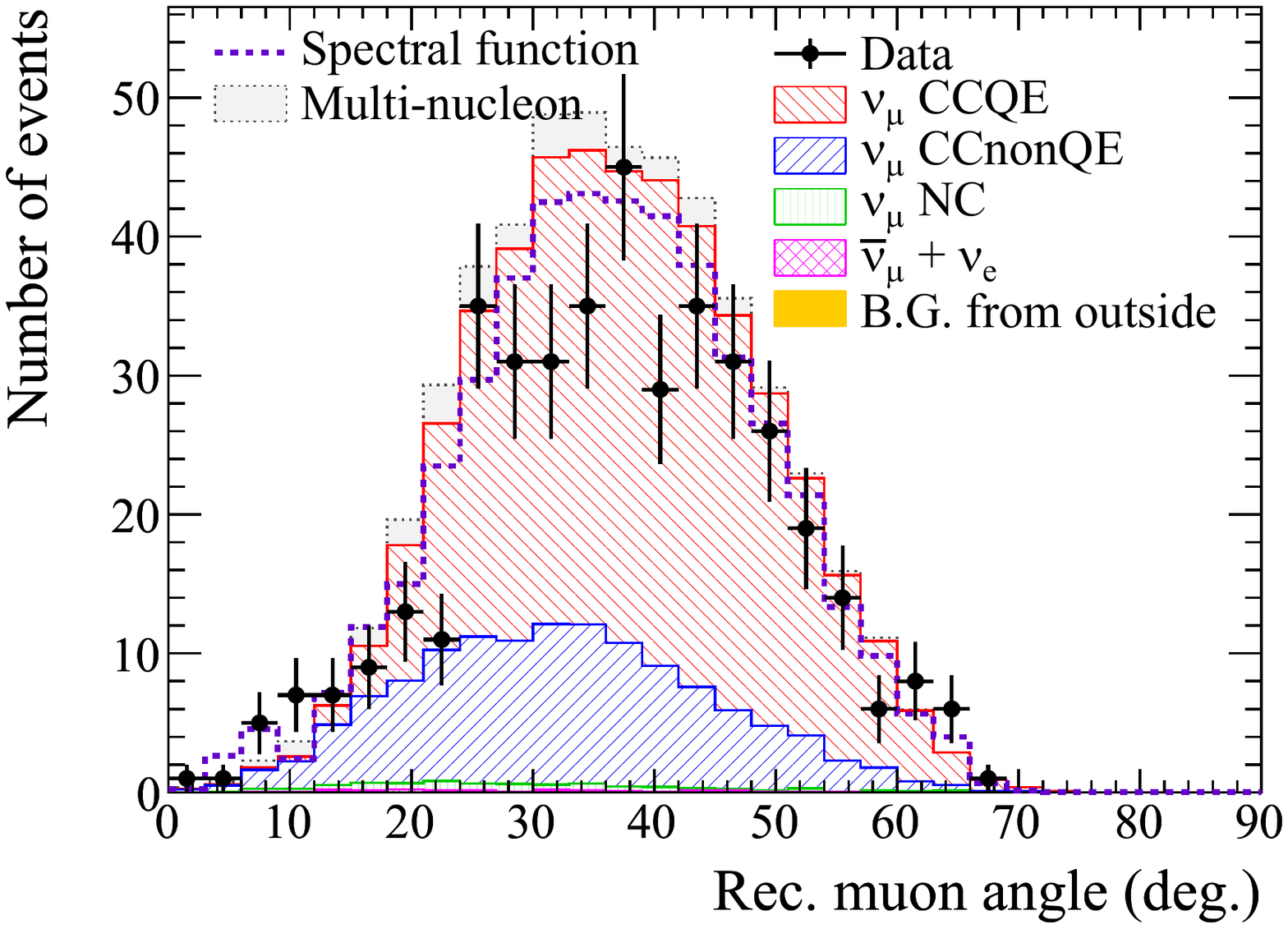}
	\subcaption{Low energy two-track sample}\label{muang_le2}
\end{minipage}
\caption{Reconstructed muon angle distribution of each CCQE candidate sample. Explanations about the predictions with the spectral function model and the multi-nucleon interaction model will be given in Secs.~\ref{sf_result} and \ref{mec_result}.}\label{ccqe_muon_angle}
\end{center}
\end{figure*}

\subsection{Results with the spectral function model}\label{sf_result}
\begin{figure}[htbp]
  \begin{center}
  \includegraphics[width=80mm]{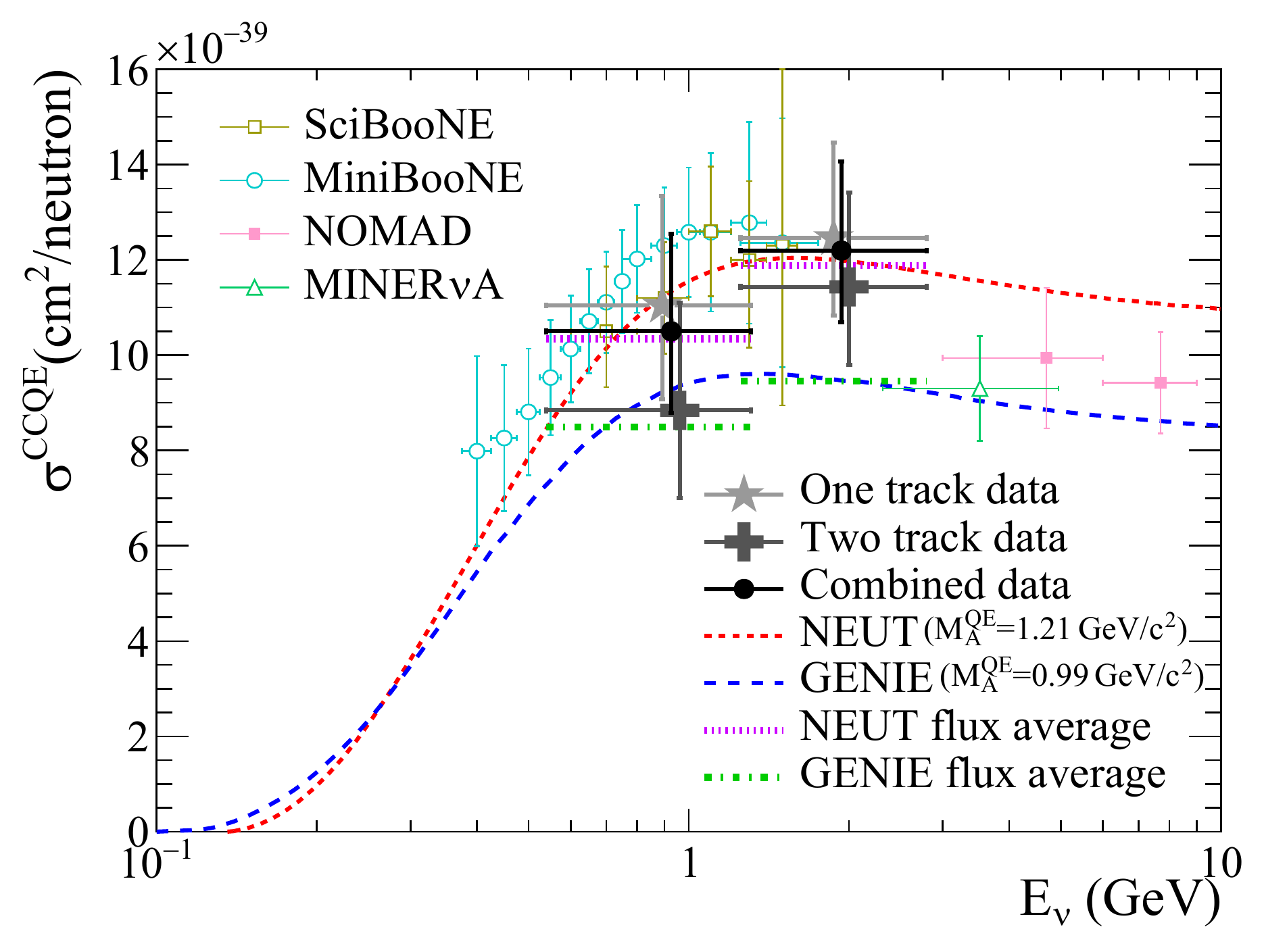}
  \caption{The CCQE cross-section results when the spectral function is used as the nuclear model.}
  \label{ccqe_result_sf}
  \end{center}
\end{figure}

The T2K default interaction model uses the relativistic Fermi gas model as the nuclear model.  
As discussed above, the spectral function model is more sophisticated and is expected to provide a better description of neutrino-nucleus interactions.
When O.\,Benhar's spectral function \cite{benhar} is used in the MC simulation for the efficiency correction instead of the relativistic Fermi gas model, the CCQE cross-section results are slightly changed, as shown in Table \ref{table_ccqe_result} (b), because of the differences in the kinematics of the final state particles which arise from the differences in the initial nucleon momentum distribution.
The cross-section results derived using the spectral function model are shown in Fig.~\ref{ccqe_result_sf} together with the model predictions.
The cross-section result in the low energy region from the one-track sample (the two-track sample) with the spectral function is 5\% lower (10\% higher) than that with the relativistic Fermi gas model.
As a result, the difference in the cross-section results between one-track and two-track samples in the low energy region becomes smaller, as shown in Table\ \ref{ccqe_1trk_2trk_ratio} in which correlations of the systematic errors between the cross-section results are correctly treated.
This change mainly comes from the differences in the final state proton kinematics between the relativistic Fermi gas model and the spectral function, which cause event migrations between the one-track sample and the two-track sample.
Therefore, this result may indicate that the spectral function is a better representation of the nuclear model than the relativistic Fermi gas model.

\begin{table*}[htbp]
\begin{center}
\caption{Ratio of the CCQE cross-section result from the one-track sample to that from the two-track in the low energy region.}
\begin{tabular}{lr}
\hline \hline
Model in MC simulation & Ratio of cross-section results\\
\hline
Relativistic Fermi gas model (T2K default model)&$1.45\pm0.09(stat.)_{-0.29}^{+0.24}(syst.)$\\
Spectral function model&$1.25\pm0.08(stat.)_{-0.26}^{+0.22}(syst.)$\\
Relativistic Fermi gas model with multi-nucleon interactions&$1.42\pm0.09(stat.)_{-0.33}^{+0.27}(syst.)$\\ \hline \hline
\end{tabular}
\label{ccqe_1trk_2trk_ratio}
  \end{center}
\end{table*}

\subsection{Results on assuming multi-nucleon interactions}\label{mec_result}
\begin{figure}[htbp]
  \begin{center}
  \includegraphics[width=80mm]{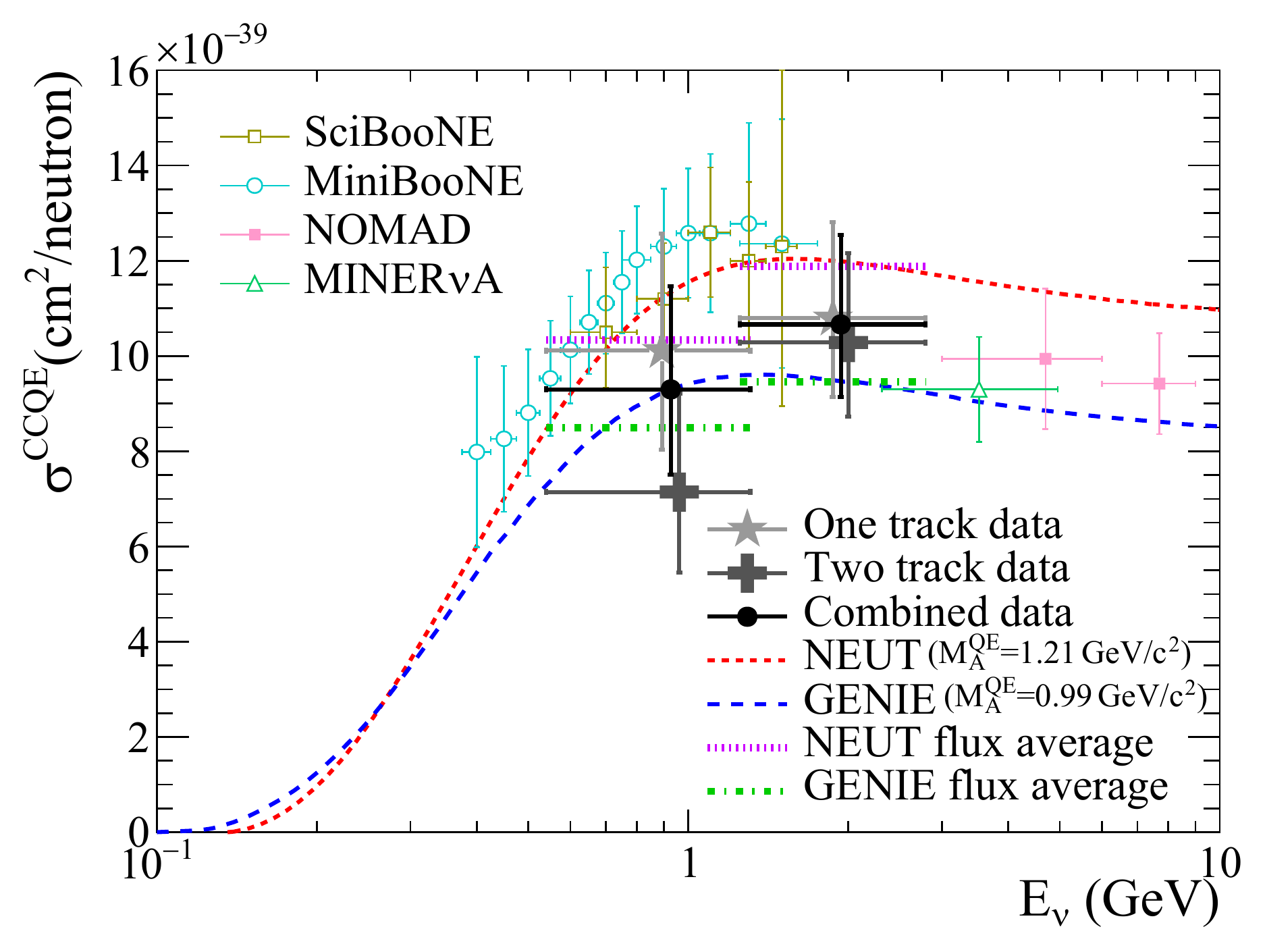}
  \caption{The CCQE cross-section results on the assumption of the existence of the multi-nucleon interaction. The relativistic Fermi gas model is used as the nuclear model.}
  \label{ccqe_result_mec}
  \end{center}
\end{figure}

The T2K default interaction model does not assume the existence of  multi-nucleon interactions via the meson exchange current.
Multi-nucleon interactions generally produce a lepton and two or more nucleons, but are misidentified as CCQE events when the additional nucleons are not reconstructed.
This means that the CCQE cross-section measurement is expected to be sensitive to the existence of multi-nucleon interactions.
Therefore, we also estimate the CCQE cross-sections assuming the existence of multi-nucleon interactions.
There are many multi-nucleon interaction models \cite{mec_martini1, mec_martini2, mec_nieves1, mec_nieves2, mec_bodek, mec_lalakulich, mec_amaro}; we used the model proposed by J. Nieves \cite{mec_nieves1, mec_nieves2}.
When the multi-nucleon interaction model was introduced into the neutrino interaction model, the CCQE model including the value of $M_A^{\mathrm{QE}}$ and the relativistic Fermi gas model were not changed in order to isolate the effect of the multi-nucleon interaction.
In this CCQE cross-section analysis, the CCQE signal is defined as the conventional two-body interaction with a single nucleon.
Therefore, the multi-nucleon interaction events are defined to be background events and are subtracted from the selected events as with CC-nonQE events, NC events, etc.
The CCQE cross-section results assuming the existence of multi-nucleon interactions are summarized in Table \ref{table_ccqe_result} (c) and shown in Fig.~\ref{ccqe_result_mec} together with the predictions.
Although the measured CCQE cross-sections assuming the existence of the multi-nucleon interaction are 6--13\% smaller, they are still compatible with the predictions.

\section{Conclusions}\label{sec:conclusion}
We have reported a CCQE cross-section measurement using the T2K on-axis neutrino detector, INGRID.
We have selected one-track and two-track samples of $\nu_\mu$ CCQE scattering in the Proton Module. From the number of selected events, the CCQE cross-sections on carbon at mean neutrino energies of 1.94~GeV and 0.93~GeV have been measured.
The cross-section analysis was performed using three different
neutrino interaction models: relativistic Fermi gas models with and without multi-nucleon interactions, and a spectral function model without multi-nucleon interactions.
Although these results are compatible with the model predictions, we found that the CCQE cross-section results are dependent on the nuclear model and the existence of multi-nucleon interactions at the 10\% level.
There is some indication, based on consistency between the one-track and two-track samples, that the event kinematics are better described by the spectral function model.
The data related to this measurement can be found electronically in \cite{data_release}.

\section*{Acknowledgments}
It is a pleasure to thank Mr.\ Taino from Mechanical Support Co.\ for
helping with the construction of INGRID.
We thank the J-PARC staff for superb accelerator performance and the
CERN NA61 collaboration for providing valuable particle production data.
We acknowledge the support of MEXT, Japan;
NSERC, NRC and CFI, Canada;
CEA and CNRS/IN2P3, France;
DFG, Germany;
INFN, Italy;
National Science Centre (NCN), Poland;
RSF, RFBR and MES, Russia;
MINECO and ERDF funds, Spain;
SNSF and SER, Switzerland;
STFC, UK; and
DOE, USA.
We also thank CERN for the UA1/NOMAD magnet,
DESY for the HERA-B magnet mover system,
NII for SINET4,
the WestGrid and SciNet consortia in Compute Canada,
GridPP, UK.
In addition participation of individual researchers
and institutions has been further supported by funds from: ERC (FP7), EU;
JSPS, Japan;
Royal Society, UK;
DOE Early Career program, USA.

\bibliographystyle{apsrev4-1}
\bibliography{main}

\providecommand{\noopsort}[1]{}\providecommand{\singleletter}[1]{#1}%
\begin{thebibliography}{72}%
\makeatletter
\providecommand \@ifxundefined [1]{%
 \@ifx{#1\undefined}
}%
\providecommand \@ifnum [1]{%
 \ifnum #1\expandafter \@firstoftwo
 \else \expandafter \@secondoftwo
 \fi
}%
\providecommand \@ifx [1]{%
 \ifx #1\expandafter \@firstoftwo
 \else \expandafter \@secondoftwo
 \fi
}%
\providecommand \natexlab [1]{#1}%
\providecommand \enquote  [1]{``#1''}%
\providecommand \bibnamefont  [1]{#1}%
\providecommand \bibfnamefont [1]{#1}%
\providecommand \citenamefont [1]{#1}%
\providecommand \href@noop [0]{\@secondoftwo}%
\providecommand \href [0]{\begingroup \@sanitize@url \@href}%
\providecommand \@href[1]{\@@startlink{#1}\@@href}%
\providecommand \@@href[1]{\endgroup#1\@@endlink}%
\providecommand \@sanitize@url [0]{\catcode `\\12\catcode `\$12\catcode
  `\&12\catcode `\#12\catcode `\^12\catcode `\_12\catcode `\%12\relax}%
\providecommand \@@startlink[1]{}%
\providecommand \@@endlink[0]{}%
\providecommand \url  [0]{\begingroup\@sanitize@url \@url }%
\providecommand \@url [1]{\endgroup\@href {#1}{\urlprefix }}%
\providecommand \urlprefix  [0]{URL }%
\providecommand \Eprint [0]{\href }%
\providecommand \doibase [0]{http://dx.doi.org/}%
\providecommand \selectlanguage [0]{\@gobble}%
\providecommand \bibinfo  [0]{\@secondoftwo}%
\providecommand \bibfield  [0]{\@secondoftwo}%
\providecommand \translation [1]{[#1]}%
\providecommand \BibitemOpen [0]{}%
\providecommand \bibitemStop [0]{}%
\providecommand \bibitemNoStop [0]{.\EOS\space}%
\providecommand \EOS [0]{\spacefactor3000\relax}%
\providecommand \BibitemShut  [1]{\csname bibitem#1\endcsname}%
\let\auto@bib@innerbib\@empty
%</preamble>
\bibitem [{\citenamefont {Abe}\ \emph {et~al.}(2011)\citenamefont {Abe} \emph
  {et~al.}}]{t2k_nim}%
  \BibitemOpen
  \bibfield  {author} {\bibinfo {author} {\bibfnamefont {K.}~\bibnamefont
  {Abe}} \emph {et~al.} (\bibinfo {collaboration} {T2K Collaboration}),\
  }\href@noop {} {\bibfield  {journal} {\bibinfo  {journal} {Nucl. Instrum.
  Meth.}\ }\textbf {\bibinfo {volume} {A659}},\ \bibinfo {pages} {106}
  (\bibinfo {year} {2011})}\BibitemShut {NoStop}%
\bibitem [{\citenamefont {Abe}\ \emph {et~al.}(2015{\natexlab{a}})\citenamefont
  {Abe} \emph {et~al.}}]{t2k_joint_oa}%
  \BibitemOpen
  \bibfield  {author} {\bibinfo {author} {\bibfnamefont {K.}~\bibnamefont
  {Abe}} \emph {et~al.} (\bibinfo {collaboration} {T2K Collaboration}),\
  }\href@noop {} {\bibfield  {journal} {\bibinfo  {journal} {arXiv:1502.01550
  [hep-ex]}\ } (\bibinfo {year} {2015}{\natexlab{a}})}\BibitemShut {NoStop}%
\bibitem [{\citenamefont {Ichikawa}(2012)}]{horn_ichikawa}%
  \BibitemOpen
  \bibfield  {author} {\bibinfo {author} {\bibfnamefont {A.~K.}\ \bibnamefont
  {Ichikawa}},\ }\href@noop {} {\bibfield  {journal} {\bibinfo  {journal}
  {Nucl. Instrum. Meth.}\ }\textbf {\bibinfo {volume} {A690}},\ \bibinfo
  {pages} {27} (\bibinfo {year} {2012})}\BibitemShut {NoStop}%
\bibitem [{\citenamefont {Abe}\ \emph {et~al.}(2012)\citenamefont {Abe} \emph
  {et~al.}}]{ingrid_nim}%
  \BibitemOpen
  \bibfield  {author} {\bibinfo {author} {\bibfnamefont {K.}~\bibnamefont
  {Abe}} \emph {et~al.} (\bibinfo {collaboration} {T2K Collaboration}),\
  }\href@noop {} {\bibfield  {journal} {\bibinfo  {journal} {Nucl. Instrum.
  Meth.}\ }\textbf {\bibinfo {volume} {A694}},\ \bibinfo {pages} {211}
  (\bibinfo {year} {2012})}\BibitemShut {NoStop}%
\bibitem [{\citenamefont {Assylbekov}\ \emph {et~al.}(2012)\citenamefont
  {Assylbekov} \emph {et~al.}}]{p0d_nim}%
  \BibitemOpen
  \bibfield  {author} {\bibinfo {author} {\bibfnamefont {S.}~\bibnamefont
  {Assylbekov}} \emph {et~al.},\ }\href@noop {} {\bibfield  {journal} {\bibinfo
   {journal} {Nucl. Instrum. Meth.}\ }\textbf {\bibinfo {volume} {A686}},\
  \bibinfo {pages} {48} (\bibinfo {year} {2012})}\BibitemShut {NoStop}%
\bibitem [{\citenamefont {Abgrall}\ \emph
  {et~al.}(2011{\natexlab{a}})\citenamefont {Abgrall} \emph
  {et~al.}}]{tpc_abgrall}%
  \BibitemOpen
  \bibfield  {author} {\bibinfo {author} {\bibfnamefont {N.}~\bibnamefont
  {Abgrall}} \emph {et~al.},\ }\href@noop {} {\bibfield  {journal} {\bibinfo
  {journal} {Nucl. Instrum. Meth.}\ }\textbf {\bibinfo {volume} {A637}},\
  \bibinfo {pages} {25} (\bibinfo {year} {2011}{\natexlab{a}})}\BibitemShut
  {NoStop}%
\bibitem [{\citenamefont {Amaudruz}\ \emph {et~al.}(2012)\citenamefont
  {Amaudruz} \emph {et~al.}}]{fgd_amaudruz}%
  \BibitemOpen
  \bibfield  {author} {\bibinfo {author} {\bibfnamefont {P.~A.}\ \bibnamefont
  {Amaudruz}} \emph {et~al.},\ }\href@noop {} {\bibfield  {journal} {\bibinfo
  {journal} {Nucl. Instrum. Meth.}\ }\textbf {\bibinfo {volume} {A696}},\
  \bibinfo {pages} {1} (\bibinfo {year} {2012})}\BibitemShut {NoStop}%
\bibitem [{\citenamefont {Allan}\ \emph {et~al.}(2013)\citenamefont {Allan}
  \emph {et~al.}}]{ecal_jinst}%
  \BibitemOpen
  \bibfield  {author} {\bibinfo {author} {\bibfnamefont {D.}~\bibnamefont
  {Allan}} \emph {et~al.},\ }\href@noop {} {\bibfield  {journal} {\bibinfo
  {journal} {Journal of Instrum.}\ }\textbf {\bibinfo {volume} {8}},\ \bibinfo
  {pages} {10019} (\bibinfo {year} {2013})}\BibitemShut {NoStop}%
\bibitem [{\citenamefont {Aoki}\ \emph {et~al.}(2013)\citenamefont {Aoki} \emph
  {et~al.}}]{smrd_nim}%
  \BibitemOpen
  \bibfield  {author} {\bibinfo {author} {\bibfnamefont {S.}~\bibnamefont
  {Aoki}} \emph {et~al.},\ }\href@noop {} {\bibfield  {journal} {\bibinfo
  {journal} {Nucl. Instrum. Meth.}\ }\textbf {\bibinfo {volume} {A698}},\
  \bibinfo {pages} {135} (\bibinfo {year} {2013})}\BibitemShut {NoStop}%
\bibitem [{\citenamefont {Fukuda}\ \emph {et~al.}(2003)\citenamefont {Fukuda}
  \emph {et~al.}}]{sk_detector}%
  \BibitemOpen
  \bibfield  {author} {\bibinfo {author} {\bibfnamefont {Y.}~\bibnamefont
  {Fukuda}} \emph {et~al.} (\bibinfo {collaboration} {Super-Kamiokande
  Collaboration}),\ }\href@noop {} {\bibfield  {journal} {\bibinfo  {journal}
  {Nucl. Instrum. Meth.}\ }\textbf {\bibinfo {volume} {A501}},\ \bibinfo
  {pages} {418} (\bibinfo {year} {2003})}\BibitemShut {NoStop}%
\bibitem [{\citenamefont {Aguilar-Arevalo}\ \emph
  {et~al.}(2010{\natexlab{a}})\citenamefont {Aguilar-Arevalo} \emph
  {et~al.}}]{miniboone_ccqe}%
  \BibitemOpen
  \bibfield  {author} {\bibinfo {author} {\bibfnamefont {A.~A.}\ \bibnamefont
  {Aguilar-Arevalo}} \emph {et~al.} (\bibinfo {collaboration} {MiniBooNE
  Collaboration}),\ }\href@noop {} {\bibfield  {journal} {\bibinfo  {journal}
  {Phys. Rev.}\ }\textbf {\bibinfo {volume} {D81}},\ \bibinfo {pages} {092005}
  (\bibinfo {year} {2010}{\natexlab{a}})}\BibitemShut {NoStop}%
\bibitem [{\citenamefont {Alcaraz-Aunion}\ and\ \citenamefont
  {Walding}(2009)}]{sciboone_ccqe}%
  \BibitemOpen
  \bibfield  {author} {\bibinfo {author} {\bibfnamefont {J.~L.}\ \bibnamefont
  {Alcaraz-Aunion}}\ and\ \bibinfo {author} {\bibfnamefont {J.}~\bibnamefont
  {Walding}} (\bibinfo {collaboration} {SciBooNE Collaboration}),\ }\href@noop
  {} {\bibfield  {journal} {\bibinfo  {journal} {AIP Conf. Proc.}\ }\textbf
  {\bibinfo {volume} {1189}},\ \bibinfo {pages} {145} (\bibinfo {year}
  {2009})}\BibitemShut {NoStop}%
\bibitem [{\citenamefont {Lyubushkin}\ \emph {et~al.}(2009)\citenamefont
  {Lyubushkin} \emph {et~al.}}]{nomad_ccqe}%
  \BibitemOpen
  \bibfield  {author} {\bibinfo {author} {\bibfnamefont {V.}~\bibnamefont
  {Lyubushkin}} \emph {et~al.} (\bibinfo {collaboration} {NOMAD
  Collaboration}),\ }\href@noop {} {\bibfield  {journal} {\bibinfo  {journal}
  {Eur. Phys. J.}\ }\textbf {\bibinfo {volume} {C63}},\ \bibinfo {pages} {355}
  (\bibinfo {year} {2009})}\BibitemShut {NoStop}%
\bibitem [{\citenamefont {Fiorentini}\ \emph {et~al.}(2013)\citenamefont
  {Fiorentini}, \citenamefont {Schmitz}, \citenamefont {Rodrigues} \emph
  {et~al.}}]{minerva_ccqe_nu}%
  \BibitemOpen
  \bibfield  {author} {\bibinfo {author} {\bibfnamefont {G.~A.}\ \bibnamefont
  {Fiorentini}}, \bibinfo {author} {\bibfnamefont {D.~W.}\ \bibnamefont
  {Schmitz}}, \bibinfo {author} {\bibfnamefont {P.~A.}\ \bibnamefont
  {Rodrigues}},  \emph {et~al.} (\bibinfo {collaboration} {MINER$\nu$A
  Collaboration}),\ }\href@noop {} {\bibfield  {journal} {\bibinfo  {journal}
  {Phys. Rev. Lett.}\ }\textbf {\bibinfo {volume} {111}},\ \bibinfo {pages}
  {022502} (\bibinfo {year} {2013})}\BibitemShut {NoStop}%
\bibitem [{\citenamefont {Auerbach}\ \emph {et~al.}(tion)\citenamefont
  {Auerbach} \emph {et~al.}}]{lsnd_ccqe}%
  \BibitemOpen
  \bibfield  {author} {\bibinfo {author} {\bibfnamefont {L.~B.}\ \bibnamefont
  {Auerbach}} \emph {et~al.} (\bibinfo {collaboration} {2002}),\ }\href@noop {}
  {\bibfield  {journal} {\bibinfo  {journal} {Phys. Rev.}\ }\textbf {\bibinfo
  {volume} {C66}},\ \bibinfo {pages} {015501} (\bibinfo {year} {LSND
  Collaboration})}\BibitemShut {NoStop}%
\bibitem [{\citenamefont {Smith}(1972)}]{smith}%
  \BibitemOpen
  \bibfield  {author} {\bibinfo {author} {\bibfnamefont {C.~H.~L.}\
  \bibnamefont {Smith}},\ }\href@noop {} {\bibfield  {journal} {\bibinfo
  {journal} {Phys. Rept.}\ }\textbf {\bibinfo {volume} {3}},\ \bibinfo {pages}
  {261} (\bibinfo {year} {1972})}\BibitemShut {NoStop}%
\bibitem [{\citenamefont {Smith}\ and\ \citenamefont {Moniz}(1972)}]{moniz}%
  \BibitemOpen
  \bibfield  {author} {\bibinfo {author} {\bibfnamefont {R.~A.}\ \bibnamefont
  {Smith}}\ and\ \bibinfo {author} {\bibfnamefont {E.~J.}\ \bibnamefont
  {Moniz}},\ }\href@noop {} {\bibfield  {journal} {\bibinfo  {journal} {Nucl.
  Phys.}\ }\textbf {\bibinfo {volume} {B43}},\ \bibinfo {pages} {605} (\bibinfo
  {year} {1972})}\BibitemShut {NoStop}%
\bibitem [{\citenamefont {Martini}\ \emph {et~al.}(2009)\citenamefont
  {Martini}, \citenamefont {Ericson}, \citenamefont {Chanfray},\ and\
  \citenamefont {Marteau}}]{mec_martini1}%
  \BibitemOpen
  \bibfield  {author} {\bibinfo {author} {\bibfnamefont {M.}~\bibnamefont
  {Martini}}, \bibinfo {author} {\bibfnamefont {M.}~\bibnamefont {Ericson}},
  \bibinfo {author} {\bibfnamefont {G.}~\bibnamefont {Chanfray}}, \ and\
  \bibinfo {author} {\bibfnamefont {J.}~\bibnamefont {Marteau}},\ }\href@noop
  {} {\bibfield  {journal} {\bibinfo  {journal} {Phys. Rev.}\ }\textbf
  {\bibinfo {volume} {C80}},\ \bibinfo {pages} {065501} (\bibinfo {year}
  {2009})}\BibitemShut {NoStop}%
\bibitem [{\citenamefont {Martini}\ \emph {et~al.}(2011)\citenamefont
  {Martini}, \citenamefont {Ericson},\ and\ \citenamefont
  {Chanfray}}]{mec_martini2}%
  \BibitemOpen
  \bibfield  {author} {\bibinfo {author} {\bibfnamefont {M.}~\bibnamefont
  {Martini}}, \bibinfo {author} {\bibfnamefont {M.}~\bibnamefont {Ericson}}, \
  and\ \bibinfo {author} {\bibfnamefont {G.}~\bibnamefont {Chanfray}},\
  }\href@noop {} {\bibfield  {journal} {\bibinfo  {journal} {Phys. Rev.}\
  }\textbf {\bibinfo {volume} {C84}},\ \bibinfo {pages} {055502} (\bibinfo
  {year} {2011})}\BibitemShut {NoStop}%
\bibitem [{\citenamefont {Nieves}\ \emph {et~al.}(2011)\citenamefont {Nieves},
  \citenamefont {Simo},\ and\ \citenamefont {Vacas}}]{mec_nieves1}%
  \BibitemOpen
  \bibfield  {author} {\bibinfo {author} {\bibfnamefont {J.}~\bibnamefont
  {Nieves}}, \bibinfo {author} {\bibfnamefont {I.~R.}\ \bibnamefont {Simo}}, \
  and\ \bibinfo {author} {\bibfnamefont {M.~J.~V.}\ \bibnamefont {Vacas}},\
  }\href@noop {} {\bibfield  {journal} {\bibinfo  {journal} {Phys. Rev.}\
  }\textbf {\bibinfo {volume} {C83}},\ \bibinfo {pages} {045501} (\bibinfo
  {year} {2011})}\BibitemShut {NoStop}%
\bibitem [{\citenamefont {Nieves}\ \emph {et~al.}(2012)\citenamefont {Nieves},
  \citenamefont {Simo},\ and\ \citenamefont {Vacas}}]{mec_nieves2}%
  \BibitemOpen
  \bibfield  {author} {\bibinfo {author} {\bibfnamefont {J.}~\bibnamefont
  {Nieves}}, \bibinfo {author} {\bibfnamefont {I.~R.}\ \bibnamefont {Simo}}, \
  and\ \bibinfo {author} {\bibfnamefont {M.~J.~V.}\ \bibnamefont {Vacas}},\
  }\href@noop {} {\bibfield  {journal} {\bibinfo  {journal} {Phys. Lett.}\
  }\textbf {\bibinfo {volume} {B707}},\ \bibinfo {pages} {72} (\bibinfo {year}
  {2012})}\BibitemShut {NoStop}%
\bibitem [{\citenamefont {Bodek}\ \emph {et~al.}(2011)\citenamefont {Bodek},
  \citenamefont {Budd},\ and\ \citenamefont {Christy}}]{mec_bodek}%
  \BibitemOpen
  \bibfield  {author} {\bibinfo {author} {\bibfnamefont {A.}~\bibnamefont
  {Bodek}}, \bibinfo {author} {\bibfnamefont {H.~S.}\ \bibnamefont {Budd}}, \
  and\ \bibinfo {author} {\bibfnamefont {M.~E.}\ \bibnamefont {Christy}},\
  }\href@noop {} {\bibfield  {journal} {\bibinfo  {journal} {Eur. Phys. J.}\
  }\textbf {\bibinfo {volume} {C71}},\ \bibinfo {pages} {1726} (\bibinfo {year}
  {2011})}\BibitemShut {NoStop}%
\bibitem [{\citenamefont {Lalakulich}\ \emph {et~al.}(2012)\citenamefont
  {Lalakulich}, \citenamefont {Gallmeister},\ and\ \citenamefont
  {Mosel}}]{mec_lalakulich}%
  \BibitemOpen
  \bibfield  {author} {\bibinfo {author} {\bibfnamefont {O.}~\bibnamefont
  {Lalakulich}}, \bibinfo {author} {\bibfnamefont {K.}~\bibnamefont
  {Gallmeister}}, \ and\ \bibinfo {author} {\bibfnamefont {U.}~\bibnamefont
  {Mosel}},\ }\href@noop {} {\bibfield  {journal} {\bibinfo  {journal} {Phys.
  Rev.}\ }\textbf {\bibinfo {volume} {C86}},\ \bibinfo {pages} {014614}
  (\bibinfo {year} {2012})}\BibitemShut {NoStop}%
\bibitem [{\citenamefont {Amaro}\ \emph {et~al.}(2012)\citenamefont {Amaro},
  \citenamefont {Barbaro}, \citenamefont {Caballero},\ and\ \citenamefont
  {Donnelly}}]{mec_amaro}%
  \BibitemOpen
  \bibfield  {author} {\bibinfo {author} {\bibfnamefont {J.~E.}\ \bibnamefont
  {Amaro}}, \bibinfo {author} {\bibfnamefont {M.~B.}\ \bibnamefont {Barbaro}},
  \bibinfo {author} {\bibfnamefont {J.~A.}\ \bibnamefont {Caballero}}, \ and\
  \bibinfo {author} {\bibfnamefont {T.~W.}\ \bibnamefont {Donnelly}},\
  }\href@noop {} {\bibfield  {journal} {\bibinfo  {journal} {Phys. Rev. Lett.}\
  }\textbf {\bibinfo {volume} {108}},\ \bibinfo {pages} {152501} (\bibinfo
  {year} {2012})}\BibitemShut {NoStop}%
\bibitem [{\citenamefont {Ankowski}(2012)}]{ankowski}%
  \BibitemOpen
  \bibfield  {author} {\bibinfo {author} {\bibfnamefont {A.~M.}\ \bibnamefont
  {Ankowski}},\ }\href@noop {} {\bibfield  {journal} {\bibinfo  {journal}
  {Phys. Rev.}\ }\textbf {\bibinfo {volume} {C86}},\ \bibinfo {pages} {024616}
  (\bibinfo {year} {2012})}\BibitemShut {NoStop}%
\bibitem [{\citenamefont {Hayato}(2009)}]{neut_hayato}%
  \BibitemOpen
  \bibfield  {author} {\bibinfo {author} {\bibfnamefont {Y.}~\bibnamefont
  {Hayato}},\ }\href@noop {} {\bibfield  {journal} {\bibinfo  {journal} {Acta
  Phys. Polon.}\ }\textbf {\bibinfo {volume} {B40}},\ \bibinfo {pages} {2477}
  (\bibinfo {year} {2009})}\BibitemShut {NoStop}%
\bibitem [{\citenamefont {Yokoyama}\ \emph {et~al.}(2009)\citenamefont
  {Yokoyama} \emph {et~al.}}]{mppc_yokoyama1}%
  \BibitemOpen
  \bibfield  {author} {\bibinfo {author} {\bibfnamefont {M.}~\bibnamefont
  {Yokoyama}} \emph {et~al.},\ }\href@noop {} {\bibfield  {journal} {\bibinfo
  {journal} {Nucl. Instrum. Meth.}\ }\textbf {\bibinfo {volume} {A610}},\
  \bibinfo {pages} {128} (\bibinfo {year} {2009})}\BibitemShut {NoStop}%
\bibitem [{\citenamefont {Yokoyama}\ \emph {et~al.}(2010)\citenamefont
  {Yokoyama} \emph {et~al.}}]{mppc_yokoyama2}%
  \BibitemOpen
  \bibfield  {author} {\bibinfo {author} {\bibfnamefont {M.}~\bibnamefont
  {Yokoyama}} \emph {et~al.},\ }\href@noop {} {\bibfield  {journal} {\bibinfo
  {journal} {Nucl. Instrum. Meth.}\ }\textbf {\bibinfo {volume} {A622}},\
  \bibinfo {pages} {567} (\bibinfo {year} {2010})}\BibitemShut {NoStop}%
\bibitem [{\citenamefont {Vacheret}\ \emph {et~al.}(2007)\citenamefont
  {Vacheret} \emph {et~al.}}]{tfb}%
  \BibitemOpen
  \bibfield  {author} {\bibinfo {author} {\bibfnamefont {A.}~\bibnamefont
  {Vacheret}} \emph {et~al.},\ }\href@noop {} {}\bibinfo {howpublished} {in
  Nuclear Science Symposium Conference Record, 2007. NSS f07. IEEE 3}
  (\bibinfo {year} {2007})\BibitemShut {NoStop}%
\bibitem [{\citenamefont {Abe}\ \emph {et~al.}(2013{\natexlab{a}})\citenamefont
  {Abe} \emph {et~al.}}]{flux_prediction}%
  \BibitemOpen
  \bibfield  {author} {\bibinfo {author} {\bibfnamefont {K.}~\bibnamefont
  {Abe}} \emph {et~al.} (\bibinfo {collaboration} {T2K Collaboration}),\
  }\href@noop {} {\bibfield  {journal} {\bibinfo  {journal} {Phys. Rev.}\
  }\textbf {\bibinfo {volume} {D87}},\ \bibinfo {pages} {012001} (\bibinfo
  {year} {2013}{\natexlab{a}})}\BibitemShut {NoStop}%
\bibitem [{\citenamefont {Brun}\ \emph {et~al.}(1993)\citenamefont {Brun} \emph
  {et~al.}}]{geant3}%
  \BibitemOpen
  \bibfield  {author} {\bibinfo {author} {\bibfnamefont {R.}~\bibnamefont
  {Brun}} \emph {et~al.},\ }\href@noop {} {}\bibinfo {howpublished} {Cern
  Program Library Long Write-up W5013} (\bibinfo {year} {1993})\BibitemShut
  {NoStop}%
\bibitem [{\citenamefont {Ferrari}\ \emph {et~al.}(2005)\citenamefont {Ferrari}
  \emph {et~al.}}]{fluka1}%
  \BibitemOpen
  \bibfield  {author} {\bibinfo {author} {\bibfnamefont {A.}~\bibnamefont
  {Ferrari}} \emph {et~al.},\ }\href@noop {} {}\bibinfo {howpublished}
  {CERN-2005-010, SLAC-R-773 and INFN-TC-05-011} (\bibinfo {year}
  {2005})\BibitemShut {NoStop}%
\bibitem [{\citenamefont {Battistoni}\ \emph {et~al.}(2007)\citenamefont
  {Battistoni} \emph {et~al.}}]{fluka2}%
  \BibitemOpen
  \bibfield  {author} {\bibinfo {author} {\bibfnamefont {G.}~\bibnamefont
  {Battistoni}} \emph {et~al.},\ }\href@noop {} {\bibfield  {journal} {\bibinfo
   {journal} {AIP Conf. Proc.}\ }\textbf {\bibinfo {volume} {896}},\ \bibinfo
  {pages} {31} (\bibinfo {year} {2007})}\BibitemShut {NoStop}%
\bibitem [{\citenamefont {Zeitnitz}\ and\ \citenamefont
  {Gabriel}(1993)}]{gcalor}%
  \BibitemOpen
  \bibfield  {author} {\bibinfo {author} {\bibfnamefont {C.}~\bibnamefont
  {Zeitnitz}}\ and\ \bibinfo {author} {\bibfnamefont {T.~A.}\ \bibnamefont
  {Gabriel}},\ }\href@noop {} {}\bibinfo {howpublished} {In Proc. of
  International Conference on Calorimetry in High Energy Physics} (\bibinfo
  {year} {1993})\BibitemShut {NoStop}%
\bibitem [{\citenamefont {Abgrall}\ \emph
  {et~al.}(2011{\natexlab{b}})\citenamefont {Abgrall} \emph {et~al.}}]{na61_1}%
  \BibitemOpen
  \bibfield  {author} {\bibinfo {author} {\bibfnamefont {N.}~\bibnamefont
  {Abgrall}} \emph {et~al.},\ }\href@noop {} {\bibfield  {journal} {\bibinfo
  {journal} {Phys. Rev.}\ }\textbf {\bibinfo {volume} {C84}},\ \bibinfo {pages}
  {034604} (\bibinfo {year} {2011}{\natexlab{b}})}\BibitemShut {NoStop}%
\bibitem [{\citenamefont {Abgrall}\ \emph {et~al.}(2012)\citenamefont {Abgrall}
  \emph {et~al.}}]{na61_2}%
  \BibitemOpen
  \bibfield  {author} {\bibinfo {author} {\bibfnamefont {N.}~\bibnamefont
  {Abgrall}} \emph {et~al.},\ }\href@noop {} {\bibfield  {journal} {\bibinfo
  {journal} {Phys. Rev.}\ }\textbf {\bibinfo {volume} {C85}},\ \bibinfo {pages}
  {035210} (\bibinfo {year} {2012})}\BibitemShut {NoStop}%
\bibitem [{\citenamefont {Eichten}\ \emph {et~al.}(1972)\citenamefont {Eichten}
  \emph {et~al.}}]{eichten}%
  \BibitemOpen
  \bibfield  {author} {\bibinfo {author} {\bibfnamefont {T.}~\bibnamefont
  {Eichten}} \emph {et~al.},\ }\href@noop {} {\bibfield  {journal} {\bibinfo
  {journal} {Nucl. Phys.}\ }\textbf {\bibinfo {volume} {B44}},\ \bibinfo
  {pages} {333} (\bibinfo {year} {1972})}\BibitemShut {NoStop}%
\bibitem [{\citenamefont {Allaby}\ \emph {et~al.}(1970)\citenamefont {Allaby}
  \emph {et~al.}}]{allaby}%
  \BibitemOpen
  \bibfield  {author} {\bibinfo {author} {\bibfnamefont {J.}~\bibnamefont
  {Allaby}} \emph {et~al.},\ }\href@noop {} {}\bibinfo {howpublished}
  {ERN-70-12} (\bibinfo {year} {1970})\BibitemShut {NoStop}%
\bibitem [{\citenamefont {Rein}\ and\ \citenamefont
  {Sehgal}(1981)}]{rands_res}%
  \BibitemOpen
  \bibfield  {author} {\bibinfo {author} {\bibfnamefont {D.}~\bibnamefont
  {Rein}}\ and\ \bibinfo {author} {\bibfnamefont {L.~M.}\ \bibnamefont
  {Sehgal}},\ }\href@noop {} {\bibfield  {journal} {\bibinfo  {journal} {Ann.
  Phys.}\ }\textbf {\bibinfo {volume} {133}},\ \bibinfo {pages} {79} (\bibinfo
  {year} {1981})}\BibitemShut {NoStop}%
\bibitem [{\citenamefont {Rein}\ and\ \citenamefont
  {Sehgal}(1983)}]{rands_coh}%
  \BibitemOpen
  \bibfield  {author} {\bibinfo {author} {\bibfnamefont {D.}~\bibnamefont
  {Rein}}\ and\ \bibinfo {author} {\bibfnamefont {L.~M.}\ \bibnamefont
  {Sehgal}},\ }\href@noop {} {\bibfield  {journal} {\bibinfo  {journal} {Nucl.
  Phys.}\ }\textbf {\bibinfo {volume} {B223}},\ \bibinfo {pages} {29} (\bibinfo
  {year} {1983})}\BibitemShut {NoStop}%
\bibitem [{\citenamefont {Gluck}\ \emph {et~al.}(1998)\citenamefont {Gluck},
  \citenamefont {Reya},\ and\ \citenamefont {Vogt}}]{gluck}%
  \BibitemOpen
  \bibfield  {author} {\bibinfo {author} {\bibfnamefont {M.}~\bibnamefont
  {Gluck}}, \bibinfo {author} {\bibfnamefont {E.}~\bibnamefont {Reya}}, \ and\
  \bibinfo {author} {\bibfnamefont {A.}~\bibnamefont {Vogt}},\ }\href@noop {}
  {\bibfield  {journal} {\bibinfo  {journal} {Eur. Phys. J.}\ }\textbf
  {\bibinfo {volume} {C5}},\ \bibinfo {pages} {461} (\bibinfo {year}
  {1998})}\BibitemShut {NoStop}%
\bibitem [{\citenamefont {Bodek}\ \emph {et~al.}(2005)\citenamefont {Bodek},
  \citenamefont {Park},\ and\ \citenamefont {Yang}}]{bodek}%
  \BibitemOpen
  \bibfield  {author} {\bibinfo {author} {\bibfnamefont {A.}~\bibnamefont
  {Bodek}}, \bibinfo {author} {\bibfnamefont {I.}~\bibnamefont {Park}}, \ and\
  \bibinfo {author} {\bibfnamefont {U.}~\bibnamefont {Yang}},\ }\href@noop {}
  {\bibfield  {journal} {\bibinfo  {journal} {Nucl. Phys. B, Proc. Suppl.}\
  }\textbf {\bibinfo {volume} {139}},\ \bibinfo {pages} {113} (\bibinfo {year}
  {2005})}\BibitemShut {NoStop}%
\bibitem [{\citenamefont {Yang}\ \emph {et~al.}(2009)\citenamefont {Yang} \emph
  {et~al.}}]{yang}%
  \BibitemOpen
  \bibfield  {author} {\bibinfo {author} {\bibfnamefont {T.}~\bibnamefont
  {Yang}} \emph {et~al.},\ }\href@noop {} {\bibfield  {journal} {\bibinfo
  {journal} {Eur. Phys. J.}\ }\textbf {\bibinfo {volume} {C63}},\ \bibinfo
  {pages} {1} (\bibinfo {year} {2009})}\BibitemShut {NoStop}%
\bibitem [{\citenamefont {Bertini}(1972)}]{bertini}%
  \BibitemOpen
  \bibfield  {author} {\bibinfo {author} {\bibfnamefont {H.~W.}\ \bibnamefont
  {Bertini}},\ }\href@noop {} {\bibfield  {journal} {\bibinfo  {journal} {Phys.
  Rev.}\ }\textbf {\bibinfo {volume} {C6}},\ \bibinfo {pages} {631} (\bibinfo
  {year} {1972})}\BibitemShut {NoStop}%
\bibitem [{\citenamefont {Lindenbaum}\ and\ \citenamefont
  {Sternheimer}(1957)}]{lindenbaum}%
  \BibitemOpen
  \bibfield  {author} {\bibinfo {author} {\bibfnamefont {S.~J.}\ \bibnamefont
  {Lindenbaum}}\ and\ \bibinfo {author} {\bibfnamefont {R.~M.}\ \bibnamefont
  {Sternheimer}},\ }\href@noop {} {\bibfield  {journal} {\bibinfo  {journal}
  {Phys. Rev.}\ }\textbf {\bibinfo {volume} {105}},\ \bibinfo {pages} {1874}
  (\bibinfo {year} {1957})}\BibitemShut {NoStop}%
\bibitem [{\citenamefont {Rowntree}\ \emph {et~al.}(1999)\citenamefont
  {Rowntree} \emph {et~al.}}]{pdd_1}%
  \BibitemOpen
  \bibfield  {author} {\bibinfo {author} {\bibfnamefont {D.}~\bibnamefont
  {Rowntree}} \emph {et~al.},\ }\href@noop {} {\bibfield  {journal} {\bibinfo
  {journal} {Phys. Rev.}\ }\textbf {\bibinfo {volume} {C60}},\ \bibinfo {pages}
  {054610} (\bibinfo {year} {1999})}\BibitemShut {NoStop}%
\bibitem [{\citenamefont {Ritchie}(1991)}]{pdd_2}%
  \BibitemOpen
  \bibfield  {author} {\bibinfo {author} {\bibfnamefont {B.~G.}\ \bibnamefont
  {Ritchie}},\ }\href@noop {} {\bibfield  {journal} {\bibinfo  {journal} {Phys.
  Rev.}\ }\textbf {\bibinfo {volume} {C44}},\ \bibinfo {pages} {533} (\bibinfo
  {year} {1991})}\BibitemShut {NoStop}%
\bibitem [{\citenamefont {Abe}\ \emph {et~al.}(2013{\natexlab{b}})\citenamefont
  {Abe} \emph {et~al.}}]{t2k_ccinc}%
  \BibitemOpen
  \bibfield  {author} {\bibinfo {author} {\bibfnamefont {K.}~\bibnamefont
  {Abe}} \emph {et~al.} (\bibinfo {collaboration} {T2K Collaboration}),\
  }\href@noop {} {\bibfield  {journal} {\bibinfo  {journal} {Phys. Rev.}\
  }\textbf {\bibinfo {volume} {D87}},\ \bibinfo {pages} {092003} (\bibinfo
  {year} {2013}{\natexlab{b}})}\BibitemShut {NoStop}%
\bibitem [{\citenamefont {Andreopoulos}\ \emph {et~al.}(2010)\citenamefont
  {Andreopoulos} \emph {et~al.}}]{genie}%
  \BibitemOpen
  \bibfield  {author} {\bibinfo {author} {\bibfnamefont {C.}~\bibnamefont
  {Andreopoulos}} \emph {et~al.},\ }\href@noop {} {\bibfield  {journal}
  {\bibinfo  {journal} {Nucl. Instrum. Meth.}\ }\textbf {\bibinfo {volume}
  {A614}},\ \bibinfo {pages} {87} (\bibinfo {year} {2010})}\BibitemShut
  {NoStop}%
\bibitem [{\citenamefont {Kuzmin}\ \emph {et~al.}(2008)\citenamefont {Kuzmin}
  \emph {et~al.}}]{m_a_ref}%
  \BibitemOpen
  \bibfield  {author} {\bibinfo {author} {\bibfnamefont {K.~S.}\ \bibnamefont
  {Kuzmin}} \emph {et~al.},\ }\href@noop {} {\bibfield  {journal} {\bibinfo
  {journal} {Eur. Phys. J.}\ }\textbf {\bibinfo {volume} {C54}},\ \bibinfo
  {pages} {517} (\bibinfo {year} {2008})}\BibitemShut {NoStop}%
\bibitem [{\citenamefont {Bodek}\ and\ \citenamefont {Ritchie}(1981)}]{bandr}%
  \BibitemOpen
  \bibfield  {author} {\bibinfo {author} {\bibfnamefont {A.}~\bibnamefont
  {Bodek}}\ and\ \bibinfo {author} {\bibfnamefont {J.~L.}\ \bibnamefont
  {Ritchie}},\ }\href@noop {} {\bibfield  {journal} {\bibinfo  {journal} {Phys.
  Rev.}\ }\textbf {\bibinfo {volume} {D24}},\ \bibinfo {pages} {1400} (\bibinfo
  {year} {1981})}\BibitemShut {NoStop}%
\bibitem [{\citenamefont {Agostinelli}\ \emph {et~al.}(2003)\citenamefont
  {Agostinelli} \emph {et~al.}}]{geant4}%
  \BibitemOpen
  \bibfield  {author} {\bibinfo {author} {\bibfnamefont {S.}~\bibnamefont
  {Agostinelli}} \emph {et~al.} (\bibinfo {collaboration} {GEANT4
  Collaboration}),\ }\href@noop {} {\bibfield  {journal} {\bibinfo  {journal}
  {Nucl. Instrum. Meth.}\ }\textbf {\bibinfo {volume} {B506}},\ \bibinfo
  {pages} {250} (\bibinfo {year} {2003})}\BibitemShut {NoStop}%
\bibitem [{\citenamefont {Apostolakis}\ \emph {et~al.}(2009)\citenamefont
  {Apostolakis} \emph {et~al.}}]{qgsp_bert}%
  \BibitemOpen
  \bibfield  {author} {\bibinfo {author} {\bibfnamefont {J.}~\bibnamefont
  {Apostolakis}} \emph {et~al.},\ }\href@noop {} {\bibfield  {journal}
  {\bibinfo  {journal} {J. Phys. Conf. Ser.}\ }\textbf {\bibinfo {volume}
  {160}},\ \bibinfo {pages} {012073} (\bibinfo {year} {2009})}\BibitemShut
  {NoStop}%
\bibitem [{\citenamefont {Birks}(1951)}]{birk1}%
  \BibitemOpen
  \bibfield  {author} {\bibinfo {author} {\bibfnamefont {J.}~\bibnamefont
  {Birks}},\ }\href@noop {} {\bibfield  {journal} {\bibinfo  {journal} {Proc.
  Phys. Soc.}\ }\textbf {\bibinfo {volume} {A64}},\ \bibinfo {pages} {874}
  (\bibinfo {year} {1951})}\BibitemShut {NoStop}%
\bibitem [{\citenamefont {Birks}(1964)}]{birk2}%
  \BibitemOpen
  \bibfield  {author} {\bibinfo {author} {\bibfnamefont {J.}~\bibnamefont
  {Birks}},\ }\href@noop {} {\emph {\bibinfo {title} {Theory and Practice of
  Scintillation Counting}}}\ (\bibinfo  {publisher} {Cambridge University
  Press},\ \bibinfo {year} {1964})\BibitemShut {NoStop}%
\bibitem [{\citenamefont {Abe}\ \emph {et~al.}(2014)\citenamefont {Abe} \emph
  {et~al.}}]{ingrid_ccincl}%
  \BibitemOpen
  \bibfield  {author} {\bibinfo {author} {\bibfnamefont {K.}~\bibnamefont
  {Abe}} \emph {et~al.} (\bibinfo {collaboration} {T2K Collaboration}),\
  }\href@noop {} {\bibfield  {journal} {\bibinfo  {journal} {Phys. Rev.}\
  }\textbf {\bibinfo {volume} {D90}},\ \bibinfo {pages} {052010} (\bibinfo
  {year} {2014})}\BibitemShut {NoStop}%
\bibitem [{\citenamefont {Aguilar-Arevalo}\ \emph
  {et~al.}(2011{\natexlab{a}})\citenamefont {Aguilar-Arevalo} \emph
  {et~al.}}]{mb_ccpi0}%
  \BibitemOpen
  \bibfield  {author} {\bibinfo {author} {\bibfnamefont {A.~A.}\ \bibnamefont
  {Aguilar-Arevalo}} \emph {et~al.} (\bibinfo {collaboration} {MiniBooNE
  Collaboration}),\ }\href@noop {} {\bibfield  {journal} {\bibinfo  {journal}
  {Phys. Rev.}\ }\textbf {\bibinfo {volume} {D83}},\ \bibinfo {pages} {052009}
  (\bibinfo {year} {2011}{\natexlab{a}})}\BibitemShut {NoStop}%
\bibitem [{\citenamefont {Aguilar-Arevalo}\ \emph
  {et~al.}(2011{\natexlab{b}})\citenamefont {Aguilar-Arevalo} \emph
  {et~al.}}]{mb_ccpipm}%
  \BibitemOpen
  \bibfield  {author} {\bibinfo {author} {\bibfnamefont {A.~A.}\ \bibnamefont
  {Aguilar-Arevalo}} \emph {et~al.} (\bibinfo {collaboration} {MiniBooNE
  Collaboration}),\ }\href@noop {} {\bibfield  {journal} {\bibinfo  {journal}
  {Phys. Rev.}\ }\textbf {\bibinfo {volume} {D83}},\ \bibinfo {pages} {052007}
  (\bibinfo {year} {2011}{\natexlab{b}})}\BibitemShut {NoStop}%
\bibitem [{\citenamefont {Aguilar-Arevalo}\ \emph
  {et~al.}(2010{\natexlab{b}})\citenamefont {Aguilar-Arevalo} \emph
  {et~al.}}]{mb_ncpi0}%
  \BibitemOpen
  \bibfield  {author} {\bibinfo {author} {\bibfnamefont {A.~A.}\ \bibnamefont
  {Aguilar-Arevalo}} \emph {et~al.} (\bibinfo {collaboration} {MiniBooNE
  Collaboration}),\ }\href@noop {} {\bibfield  {journal} {\bibinfo  {journal}
  {Phys. Rev.}\ }\textbf {\bibinfo {volume} {D81}},\ \bibinfo {pages} {013005}
  (\bibinfo {year} {2010}{\natexlab{b}})}\BibitemShut {NoStop}%
\bibitem [{\citenamefont {Fields}\ \emph {et~al.}(2013)\citenamefont {Fields},
  \citenamefont {Chvojka} \emph {et~al.}}]{minerva_ccqe_nubar}%
  \BibitemOpen
  \bibfield  {author} {\bibinfo {author} {\bibfnamefont {L.}~\bibnamefont
  {Fields}}, \bibinfo {author} {\bibfnamefont {J.}~\bibnamefont {Chvojka}},
  \emph {et~al.} (\bibinfo {collaboration} {MINER$\nu$A Collaboration}),\
  }\href@noop {} {\bibfield  {journal} {\bibinfo  {journal} {Phys. Rev. Lett.}\
  }\textbf {\bibinfo {volume} {111}},\ \bibinfo {pages} {022501} (\bibinfo
  {year} {2013})}\BibitemShut {NoStop}%
\bibitem [{\citenamefont {Eberly}\ \emph {et~al.}(2014)\citenamefont {Eberly}
  \emph {et~al.}}]{minerva_cc1pi}%
  \BibitemOpen
  \bibfield  {author} {\bibinfo {author} {\bibfnamefont {B.}~\bibnamefont
  {Eberly}} \emph {et~al.} (\bibinfo {collaboration} {MINER$\nu$A
  Collaboration}),\ }\href@noop {} {}\bibinfo {howpublished}
  {FERMILAB-PUB-14-193-E} (\bibinfo {year} {2014})\BibitemShut {NoStop}%
\bibitem [{\citenamefont {Hasegawa}\ \emph {et~al.}(2005)\citenamefont
  {Hasegawa} \emph {et~al.}}]{k2k_coh}%
  \BibitemOpen
  \bibfield  {author} {\bibinfo {author} {\bibfnamefont {M.}~\bibnamefont
  {Hasegawa}} \emph {et~al.} (\bibinfo {collaboration} {K2K Collaboration}),\
  }\href@noop {} {\bibfield  {journal} {\bibinfo  {journal} {Phys. Rev. Lett.}\
  }\textbf {\bibinfo {volume} {95}},\ \bibinfo {pages} {252301} (\bibinfo
  {year} {2005})}\BibitemShut {NoStop}%
\bibitem [{\citenamefont {Hiraide}\ \emph {et~al.}(2008)\citenamefont {Hiraide}
  \emph {et~al.}}]{sciboone_coh}%
  \BibitemOpen
  \bibfield  {author} {\bibinfo {author} {\bibfnamefont {K.}~\bibnamefont
  {Hiraide}} \emph {et~al.} (\bibinfo {collaboration} {SciBooNE
  Collaboration}),\ }\href@noop {} {\bibfield  {journal} {\bibinfo  {journal}
  {Phys. Rev.}\ }\textbf {\bibinfo {volume} {D78}},\ \bibinfo {pages} {112004}
  (\bibinfo {year} {2008})}\BibitemShut {NoStop}%
\bibitem [{\citenamefont {Kurimoto}\ \emph {et~al.}(2010)\citenamefont
  {Kurimoto} \emph {et~al.}}]{sciboone_nccoh}%
  \BibitemOpen
  \bibfield  {author} {\bibinfo {author} {\bibfnamefont {Y.}~\bibnamefont
  {Kurimoto}} \emph {et~al.} (\bibinfo {collaboration} {SciBooNE
  Collaboration}),\ }\href@noop {} {\bibfield  {journal} {\bibinfo  {journal}
  {Phys. Rev.}\ }\textbf {\bibinfo {volume} {D81}},\ \bibinfo {pages} {111102}
  (\bibinfo {year} {2010})}\BibitemShut {NoStop}%
\bibitem [{\citenamefont {Adamson}\ \emph {et~al.}(2010)\citenamefont {Adamson}
  \emph {et~al.}}]{minos_ccinc}%
  \BibitemOpen
  \bibfield  {author} {\bibinfo {author} {\bibfnamefont {P.}~\bibnamefont
  {Adamson}} \emph {et~al.} (\bibinfo {collaboration} {MINOS Collaboration}),\
  }\href@noop {} {\bibfield  {journal} {\bibinfo  {journal} {Phys. Rev.}\
  }\textbf {\bibinfo {volume} {D81}},\ \bibinfo {pages} {072002} (\bibinfo
  {year} {2010})}\BibitemShut {NoStop}%
\bibitem [{\citenamefont {Jen}\ \emph {et~al.}(2014)\citenamefont {Jen} \emph
  {et~al.}}]{sf_imp}%
  \BibitemOpen
  \bibfield  {author} {\bibinfo {author} {\bibfnamefont {C.~M.}\ \bibnamefont
  {Jen}} \emph {et~al.},\ }\href@noop {} {\bibfield  {journal} {\bibinfo
  {journal} {Phys. Rev.}\ }\textbf {\bibinfo {volume} {D90}},\ \bibinfo {pages}
  {093004} (\bibinfo {year} {2014})}\BibitemShut {NoStop}%
\bibitem [{\citenamefont {Benhar}\ \emph {et~al.}(2005)\citenamefont {Benhar},
  \citenamefont {Farina}, \citenamefont {Nakamura}, \citenamefont {Sakuda},\
  and\ \citenamefont {Seki}}]{escat1}%
  \BibitemOpen
  \bibfield  {author} {\bibinfo {author} {\bibfnamefont {O.}~\bibnamefont
  {Benhar}}, \bibinfo {author} {\bibfnamefont {N.}~\bibnamefont {Farina}},
  \bibinfo {author} {\bibfnamefont {H.}~\bibnamefont {Nakamura}}, \bibinfo
  {author} {\bibfnamefont {M.}~\bibnamefont {Sakuda}}, \ and\ \bibinfo {author}
  {\bibfnamefont {R.}~\bibnamefont {Seki}},\ }\href@noop {} {\bibfield
  {journal} {\bibinfo  {journal} {Phys. Rev.}\ }\textbf {\bibinfo {volume}
  {D72}},\ \bibinfo {pages} {053005} (\bibinfo {year} {2005})}\BibitemShut
  {NoStop}%
\bibitem [{\citenamefont {Nakamura}\ \emph {et~al.}(2007)\citenamefont
  {Nakamura}, \citenamefont {Nasu}, \citenamefont {Sakuda},\ and\ \citenamefont
  {Benhar}}]{escat2}%
  \BibitemOpen
  \bibfield  {author} {\bibinfo {author} {\bibfnamefont {H.}~\bibnamefont
  {Nakamura}}, \bibinfo {author} {\bibfnamefont {T.}~\bibnamefont {Nasu}},
  \bibinfo {author} {\bibfnamefont {M.}~\bibnamefont {Sakuda}}, \ and\ \bibinfo
  {author} {\bibfnamefont {O.}~\bibnamefont {Benhar}},\ }\href@noop {}
  {\bibfield  {journal} {\bibinfo  {journal} {Phys. Rev.}\ }\textbf {\bibinfo
  {volume} {C76}},\ \bibinfo {pages} {065208} (\bibinfo {year}
  {2007})}\BibitemShut {NoStop}%
\bibitem [{\citenamefont {Abe}\ \emph {et~al.}(2013{\natexlab{c}})\citenamefont
  {Abe} \emph {et~al.}}]{t2k_nue_2013}%
  \BibitemOpen
  \bibfield  {author} {\bibinfo {author} {\bibfnamefont {K.}~\bibnamefont
  {Abe}} \emph {et~al.} (\bibinfo {collaboration} {T2K Collaboration}),\
  }\href@noop {} {\bibfield  {journal} {\bibinfo  {journal} {Phys. Rev.}\
  }\textbf {\bibinfo {volume} {D88}},\ \bibinfo {pages} {032002} (\bibinfo
  {year} {2013}{\natexlab{c}})}\BibitemShut {NoStop}%
\bibitem [{\citenamefont {Abe}\ \emph {et~al.}(2015{\natexlab{b}})\citenamefont
  {Abe} \emph {et~al.}}]{suppl_material}%
  \BibitemOpen
  \bibfield  {author} {\bibinfo {author} {\bibfnamefont {K.}~\bibnamefont
  {Abe}} \emph {et~al.},\ }\href@noop {} {\enquote {\bibinfo {title}
  {Supplemental material},}\ } (\bibinfo {year} {2015}{\natexlab{b}}),\
  \bibinfo {note}
  {\url{http://link.aps.org/supplemental/10.1103/PhysRevD.91.112002}}\BibitemShut
  {NoStop}%
\bibitem [{\citenamefont {Benhar}\ \emph {et~al.}(1994)\citenamefont {Benhar}
  \emph {et~al.}}]{benhar}%
  \BibitemOpen
  \bibfield  {author} {\bibinfo {author} {\bibfnamefont {O.}~\bibnamefont
  {Benhar}} \emph {et~al.},\ }\href@noop {} {\bibfield  {journal} {\bibinfo
  {journal} {Nucl. Phys.}\ }\textbf {\bibinfo {volume} {A579}},\ \bibinfo
  {pages} {493} (\bibinfo {year} {1994})}\BibitemShut {NoStop}%
\bibitem [{\citenamefont {Abe}\ \emph {et~al.}(2015{\natexlab{c}})\citenamefont
  {Abe} \emph {et~al.}}]{data_release}%
  \BibitemOpen
  \bibfield  {author} {\bibinfo {author} {\bibfnamefont {K.}~\bibnamefont
  {Abe}} \emph {et~al.},\ }\href@noop {} {\enquote {\bibinfo {title} {{T2K}
  public data},}\ } (\bibinfo {year} {2015}{\natexlab{c}}),\ \bibinfo {note}
  {\url{http://t2k-experiment.org/results/ingriddata-numu-ccqe-xs-2015}}\BibitemShut
  {NoStop}%
\end{thebibliography}%

\end{document}